\documentstyle[12pt,epsf]{article}
\begin{document}

\baselineskip 18pt

\newcommand{\sheptitle} {Fermion Masses in a Supersymmetric
SU(4)$\otimes$SU(2)$_L\otimes$SU(2)$_R$ Model}

\newcommand{\shepauthor} {B. C. Allanach and S. F. King }

\newcommand{\shepaddress} {Physics Department, University of
Southampton\\Southampton, SO17 5NH, U.K.}

\newcommand{\shepabstract} {We calculate quark and lepton masses and
quark mixing angles in the framework of a supersymmetric
SU(4)$\otimes$SU(2)$_L\otimes$SU(2)$_R$ model where the gauge group is
broken at 10$^{16}$ GeV.  The model predicts third family
top-bottom-tau Yukawa unification, as in minimal SO(10).  The other
smaller Yukawa couplings are assumed to arise from non-renormalisable
operators suppressed by powers of some heavy scale. We perform a
systematic operator analysis of the model in order to find the minimum
set of operators which describe the low energy quark and lepton
masses, and quark mixing angles consistent with low-energy
phenomenology.  A novel feature of the model is the possibility of
asymmetric texture zeroes in the Yukawa matrices at the scale of the
new physics.  Successful predictions are obtained for $m_t$, $\tan
\beta$, $m_s/m_\mu$, $m_d/m_e$ and $V_{ub}$ in terms of a CP violating
phase $\phi$. For example, we predict $\tan \beta = 35-65$, $m_t = 130
- 190$ GeV, and $|V_{ub}|>0.0040$.}

\begin{titlepage}
\begin{flushright} SHEP 94-05 \\
\end{flushright}
\vspace{.4in}
\begin{center} {\large{\bf \sheptitle}}
\bigskip \\ \shepauthor \\ \mbox{} \\ {\it \shepaddress} \\
\vspace{.5in} {\bf Abstract} \bigskip \end{center}
\setcounter{page}{0}
\shepabstract
\end{titlepage}

\section{Introduction}

The problem of quark and lepton masses is one of the most fascinating
and perplexing problems of particle physics.  The standard model,
despite its successes, can offer no glimpse of insight into the
apparently bizarre pattern of masses and mixing angles which
experiment has presented us with.  We do not even know why there are
three families rather than one. It is clear, then, that in order to
gain some insight into the fermion mass problem, one must go beyond
the standard model. The big question of course is what lies beyond it?

We have not yet experimentally studied the mechanism of electroweak
symmetry breaking, so one might argue that it is premature to study
the fermion mass problem. Unless we can answer this, we have no hope
of understanding anything about fermion masses since we do not have a
starting point from which to analyse the problem.  However LEP has
taught us that whatever breaks electroweak symmetry must do so in a
way which very closely resembles the standard model. This observation
by itself is enough to disfavour many dynamical models involving large
numbers of new fermions.  By contrast the minimal supersymmetric
standard model (MSSM) mimics the standard model very closely.
Furthermore, by accurately measuring the strong coupling constant, LEP
has shown that the gauge couplings of the MSSM merge very accurately
at a scale just above $10^{16}$ GeV, thus providing a hint for
possible unification at this scale. On the theoretical side,
supersymmetry (SUSY) and grand unified theories (GUTs) fit together
very nicely in several ways, providing a solution to the technical
hierarchy problem for example. When SUSY GUTs are extended to
supergravity (SUGRA) the beautiful picture of universal soft SUSY
breaking parameters and radiative electroweak symmetry breaking via a
large top quark yukawa coupling emerges. Finally, there is an on-going
effort to embed all of this structure in superstring models, thereby
allowing a complete unification, including gravity.

Given the promising scenario mentioned above, it is hardly surprising
that many authors have turned to the SUSY GUT framework as a
springboard from which to attack the problem of fermion masses
\cite{Rabyi}.
Indeed in recent years there has been a flood of papers on fermion
masses in SUSY GUTs.  Although the approaches differ in detail, there
are some common successful themes which have been known for some
time. For example the idea of bottom-tau Yukawa unification in SUSY
GUTs \cite{btauold} works well with current data
\cite{btaunew}.
A more ambitious extension of this idea is the Georgi-Jarlskog (GJ)
ansatz which provides a successful description of all down-type quark
and charged lepton masses
\cite{GJ,Yuk}, and which also works well with current data
\cite{DHR}.
The GJ approach involves the idea of texture zeroes and predicts
simple relations at the unification scale which are then evolved to
low-energies using the renormalisation group (RG) equations.  These
approaches are concerned with general properties of the mass matrices,
rather than those of specific models.

In order to understand the origin of the texture zeroes, one must
consider the details of the model above the scale $M_X \sim 10^{16}$
GeV (SO(10) for example in the case of GJ).  While one might not wish
to restrict oneself to some particular gauge symmetry at $M_X$, it is
almost essential to specify the model at this scale in order to make
any progress at all.  The alternative is to simply make a list of
assumptions about the nature of the Yukawa matrices at $M_X$
\cite{frogetal}.  For example Ramond, Roberts and Ross (RRR)
\cite{RRR} assumed symmetric Yukawa matrices at $M_X$, together with
the GJ ansatz for the lepton sector.  It is difficult to proceed
beyond this without specifying a particular model. Indeed, this model
dependence may be a good thing since it may mean that the fermion mass
spectrum at low energies is sensitive to the theory at $M_X$, so it
can be used as a window into the high-energy theory.  Therefore in
what follows we shall restrict ourselves to the very specific gauge
group at $M_X$ referred to in the title. Our motivation for
considering this particular theory is discussed below.

Twenty-one years ago Pati and Salam proposed a model in which the
standard model was embedded in the gauge group SU(4)$\otimes$SU(2)$_L
\otimes $SU(2)$_R$ \cite{pati}.  More recently a superpersymmetric
(SUSY) version of this model was proposed in which the gauge group is
broken at $M_{X}\sim 10^{16}$ GeV \cite{leo1}.  The model \cite{leo1}
does not involve adjoint representations and later some attempt was
made to derive it from four-dimensional strings, although there are
some difficulties with the current formulation \cite{leo2}.  In this
paper we shall not be concerned with the superstring formulation of
the model, but instead we shall focus on the ``low-energy'' effective
field theory.  The absence of adjoint representations is not an
essential prerequisite for the model to descend from the superstring,
but it leads to some technical simplifications.  Also in the present
model, the colour triplets which are in separate representations from
the Higgs doublets, become heavy in a very simple way so the Higgs
doublet-triplet splitting problem does not arise.  These two features
(absence of adjoint representations and absence of the doublet-triplet
splitting problem,) are shared by flipped SU(5)$\otimes $U(1)
\cite{flipped}, which also has a superstring formulation.
Although the present model and flipped SU(5)$\otimes $U(1) are similar
in many ways, there are some important differences.  Whereas the
Yukawa matrices of flipped SU(5)$\otimes$U(1) are completely unrelated
at the level of the effective field theory at $M_{X}$ (although they
may have relations coming from the string model) in the present model
there is a constraint that the top, bottom and tau Yukawa couplings
must all unify at that scale.  In addition there will be Clebsch
relations between the other elements of the Yukawa matrices, assuming
they are described by non-renormalisable operators, which would not be
present in flipped SU(5)$\otimes $U(1). In these respects the model
resembles the SO(10) model recently analysed by Anderson et al
\cite{Larry}.  However it differs from the SO(10) model in that the
present model does not have an SU(5) subgroup which is central to the
analysis of the SO(10) model. In addition the operator structure of
the present model is totally different.  Thus the model under
consideration is in some sense similar to flipped SU(5)$\otimes$U(1),
but has third family Yukawa unification and precise Clebsch
relationships as in SO(10).  We find this combination of features
quite remarkable, and it seems to us that this provides a rather
strong motivation to study the problem of fermion masses in this
model.

The problem of fermion masses in the
SU(4)$\otimes$SU(2)$_L\otimes$SU(2)$_R$ model was recently considered
by one of us \cite{422}.  It was implicitly assumed that the Yukawa
matrices were symmetric, and it was shown that by introducing suitable
operators the model could make contact with the successful RRR ansatz,
which incorporate the GJ ansatz, at the expense of fine-tuning the
coefficients of the operators \cite{422}. Essentially the small
entries in the RRR matrices were obtained by assuming that the
coefficients of two operators were tuned to partially cancel.  The
purpose of the present paper is threefold. First we shall generalise
the analysis to the case of non-symmetric Yukawa matrices, since there
is no symmetry which enforces symmetric Yukawa matrices in this
model\footnote{Even the imposition of parity does not lead to
symmetric Yukawa matrices (see Section 3.3.)}.  This allows the
possibility of asymmetric texture zeroes, which as far as we are aware
have never been considered before.  Of course this means that we
cannot rely on the RRR analysis, and therefore we perform our own
phenomenological analysis of the quark and lepton masses and quark
mixing angles.  Second we extend the operator analysis to consider
many other operators not considered in the previous analysis.  In fact
we search for all possible low dimensional operators, and
systematically search for the minimum set with which to describe the
spectrum.  Third we impose a naturalness criterion and reject all
possibilities which involve fine-tuning the coefficients of operators.
The result of all this is a small set of possible solutions to the
problem of quark and lepton masses and CKM angles in this model.

The remainder of the paper is organised as follows.  In section 2 we
briefly summarise the model.  In section 3 we describe the operator
strategy we employ.  In section 4 the details of the calculation are
outlined, including the RG and CKM analysis. The ansatze, results and
predictions are presented.  Section 5 contains our conclusions about
the previous analysis, and a brief discussion of theoretical
uncertainties involved with the calculation. In the appendices we list
the operators in explicit component form.

\section{The Model}
Here we briefly summarise the parts of the model which are relevant
for our analysis.  For a more complete discussion see \cite{leo1}.
The gauge group is,
\begin{equation}
\mbox{SU(4)}\otimes \mbox{SU(2)}_L \otimes \mbox{SU(2)}_R. \label{422}
\end{equation}
The left-handed quarks and leptons are accommodated in the following
representations,
\begin{equation}
{F^i}^{\alpha a}=(4,2,1)=
\left(\begin{array}{cccc}
u^R & u^B & u^G & \nu \\ d^R & d^B & d^G & e^-
\end{array} \right)^i
\end{equation}
\begin{equation}
{\bar{F}}_{x \alpha}^i=(\bar{4},1,\bar{2})=
\left(\begin{array}{cccc}
\bar{d}^R & \bar{d}^B & \bar{d}^G & e^+  \\
\bar{u}^R & \bar{u}^B & \bar{u}^G & \bar{\nu}
\end{array} \right)^i
\end{equation}
where $\alpha=1\ldots 4$ is an SU(4) index, $a,x=1,2$ are
SU(2)$_{L,R}$ indices, and $i=1\ldots 3$ is a family index.  The Higgs
fields are contained in the following representations,
\begin{equation}
h_{a}^x=(1,\bar{2},2)=
\left(\begin{array}{cc}
  {h_2}^+ & {h_1}^0 \\ {h_2}^0 & {h_1}^- \\
\end{array} \right) \label{h}
\end{equation}
(where $h_1$ and $h_2$ are the low energy Higgs superfields associated
with the MSSM.) The two heavy Higgs representations are
\begin{equation}
{H}^{\alpha b}=(4,1,2)=
\left(\begin{array}{cccc}
u_H^R & u_H^B & u_H^G & \nu_H \\ d_H^R & d_H^B & d_H^G & e_H^-
\end{array} \right) \label{H}
\end{equation}
and
\begin{equation}
{\bar{H}}_{\alpha x}=(\bar{4},1,\bar{2})=
\left(\begin{array}{cccc}
\bar{d}_H^R & \bar{d}_H^B & \bar{d}_H^G & e_H^+ \\
\bar{u}_H^R & \bar{u}_H^B & \bar{u}_H^G & \bar{\nu}_H
\end{array} \right). \label{barH}
\end{equation}

The Higgs fields are assumed to develop VEVs,
\begin{equation}
<H>=<\nu_H>\sim M_{X}, \ \ <\bar{H}>=<\bar{\nu}_H>\sim M_{X}
\label{HVEV}
\end{equation}
leading to the symmetry breaking at $M_{X}$
\begin{equation}
\mbox{SU(4)}\otimes \mbox{SU(2)}_L \otimes \mbox{SU(2)}_R
\longrightarrow
\mbox{SU(3)}_C \otimes \mbox{SU(2)}_L \otimes \mbox{U(1)}_Y
\label{422to321}
\end{equation}
in the usual notation.  Under the symmetry breaking in
Eq.\ref{422to321}, the Higgs field $h$ in Eq.\ref{h} splits into two
Higgs doublets $h_1$, $h_2$ whose neutral components subsequently
develop weak scale VEVs,
\begin{equation}
<h_1^0>=v_1, \ \ <h_2^0>=v_2 \label{vevs}
\end{equation}
with $\tan \beta \equiv v_2/v_1$.

In addition to the Higgs fields in Eqs.~\ref{H},\ref{barH} the model
also involves an SU(4) sextet field $D=(6,1,1)$ and three singlet
fields $\phi_m=(1,1,1)$ which do not acquire VEVs plus one singlet
field $N=(1,1,1)$ which acquires a weak scale VEV $<N>=x$. The
superpotential, suppressing gauge indices, is then
\cite{leo1}\footnote{The resulting low energy theory may resemble the
so-called Next-to-Minimal Supersymmetric Standard Model (NMSSM)
involving an extra gauge singlet. However for simplicity our
calculations will be based on the MSSM.}
\begin{equation}
W=\lambda^{ij}_1F_i\bar{F}_jh+\lambda^{im}_{2}\bar{F}_iH\phi_m
+\lambda_3HHD+\lambda_4\bar{H}\bar{H}D+\lambda_5hhN
+\lambda_6^{mnq}\phi_m\phi_n\phi_q +\lambda_7N^3 \label{W}
\end{equation}
Note that this is not the most general superpotential invariant under
the gauge symmetry. Additional terms not included in Eq.\ref{W} may be
forbidden by imposing suitable discrete symmetries, the details of
which need not concern us here.  The $D$ field does not develop a VEV
but the terms in Eq.\ref{W} $HHD$ and $\bar{H} \bar{H}D$ combine the
colour triplet parts of $H$, $\bar{H}$ and $D$ into acceptable
GUT-scale mass terms \cite{leo1}. The $\phi_m$ fields play an
important part in ensuring that the right-handed neutrinos gain large
masses, leading to acceptably small observable neutrino masses.  The
effect depends on terms in the superpotential like $\bar{F}H\phi$ and
$\phi^3$ \cite{twoloop}.  Below $M_{X}$ the part of the superpotential
involving quark and charged lepton fields is just
\begin{equation}
W =\lambda^{ij}_UQ_i\bar{U}_jh_2+\lambda^{ij}_DQ_i\bar{D}_jh_1
+\lambda^{ij}_EL_i\bar{E}_jh_1+ \ldots \label{NMSSM}
\end{equation}
with the boundary conditions at $M_{X}$,
\begin{equation}
\lambda^{ij}_1=\lambda^{ij}_U=\lambda^{ij}_D=\lambda^{ij}_E,
\ \ \lambda_5=\lambda, \ \ \lambda_7=k \label{boundary}
\end{equation}

The model just described must explain why the gauge couplings which
are roughly equal at $M_{X}\sim 10^{16}$ GeV remain roughly equal up
to the compactification scale $M_{c}\sim 10^{17}$ GeV.  Conventional
SUSY GUTs achieve this in the most direct way possible, by embedding
the standard gauge group into a simple gauge group with a single gauge
coupling constant. However, conventional SUSY GUTs are not fully
unified because they do not include gravity. The only consistent known
theories of gravity are string theories, and string theories which
allow adjoint superfields are quite cumbersome \cite{string}. On the
other hand string theories that do not involve adjoint superfields,
and consequently cannot involve a simple gauge group, must explain why
the gauge couplings which appear to be unified at $M_{X}$ are in fact
unified at $M_{c}$.

Recently it was suggested by one of us \cite{422} that an attractive
solution to this problem is to introduce some additional GUT-scale
superfields in order to make the model left-right symmetric,
\begin{equation}
H'=(4,2,1), \ \ \bar{H}'=(\bar{4},\bar{2},1) \label{H'}
\end{equation}
Having guaranteed the equality of the SU(2)$_{L,R}$ couplings
$g_L=g_R$, it is possible to require that the SU(4) beta function,
$\beta_4$, is equal to the common SU(2) beta functions, $\beta_2$, so
that if the gauge couplings are equal at $M_{X}$ then they will remain
equal above this scale.  The one-loop SUSY $\beta$ functions are,
\begin{equation}
\beta_i=\mu\frac{\partial \alpha_i}{\partial \mu}=
-\frac{b_i}{2\pi}\alpha_i^2+\ldots
\end{equation}
where in the model defined in the previous section, and augmented by
the Higgs representations in Eq.\ref{H'} we find
\begin{equation}
b_4=[6-n_D-4n_H], \ \ b_2=[-1-4n_H] \label{beta}
\end{equation}
where we have allowed for $n_D$ copies of the sextet superfield $D$,
and $n_H$ copies of the set of fields $(H,\bar{H},H',\bar{H}')$.
{}From Eqs.~\ref{H'},~\ref{beta} it is clear that the combination of
left-right symmetry and the choice $n_D=7$ (for any choice of $n_H$)
is sufficient to guarantee that if the couplings are equal at $M_{X}$
then they will remain equal above this scale to one-loop order,
ignoring threshold effects.  However, as we shall see in
section~\ref{sec:parity}, such a left-right symmetry does not lead to
any simplifications of the Yukawa matrices at $M_X$, and so we shall
not impose such a symmetry in this paper.

\section{Operators}

\subsection{The Basic Strategy}

In this model the two Higgs doublets are unified into a single
representation $h$ in Eq.\ref{h} and this leads to the GUT-scale
equality of the three Yukawa matrices in
Eqs.\ref{NMSSM},~\ref{boundary}. This boundary condition also applies
to the version of the conventional SUSY GUT based on SO(10) in which
both Higgs doublets are unified into a single 10 representation.  As
it turns out, the idea of Yukawa unification works rather well for the
third family \cite{Yuk}, leading to the prediction of a large top
quark mass $m_t>165$ GeV, and $\tan \beta \sim m_t/m_b$ where $m_b$ is
the bottom quark mass.  However Yukawa unification for the first two
families is not successful, since it would lead to unacceptable mass
relations amongst the lighter fermions, and zero mixing angles at
$M_{X}$. In the SO(10) SUSY GUT there are various ways out of these
difficulties, and if the present model is to be regarded as a
surrogate SUSY GUT it must also resolve them.

One interesting proposal has recently been put forward to account for
the fermion masses in an SO(10) SUSY GUT with a single Higgs in the 10
representation \cite{Larry}. According this approach, only the third
family is allowed to receive mass from the renormalisable operators in
the superpotential.  The remaining masses and mixings are generated
from a minimal set of just three specially chosen non-renormalisable
operators whose coefficients are suppressed by some large scale.
Furthermore these operators are only allowed to contain adjoint 45
Higgs representations, chosen from a set of fields denoted $45_Y$,
$45_{B-L}$, $45_{T_{3R}}$, $45_X$ whose VEVs point in the direction of
the generators specified by the subscripts, in the notation of
\cite{Larry}.

This is precisely the strategy we wish to follow here.  We shall
assume that only the third family receives its mass from a
renormalisable Yukawa coupling. All the other renormalisable Yukawa
couplings are set to zero. Then non-renormalisable operators are
written down which will play the role of small effective Yukawa
couplings. The effective Yukawa couplings are small because they
originate from non-renormalisable operators which are suppressed by
powers of the heavy scale $M$. In this paper we shall restrict
ourselves to all possible non-renormalisable operators which can be
constructed from different group theoretical contractions of the
fields:
\begin{equation}
O_{ij}\sim (F_i\bar{F}_j )h\left(\frac{H\bar{H}}{M^2}\right)^n+{\mbox
h.c.} \label{op}
\end{equation}
where we have used the fields $H,\bar{H}$ in Eqs.\ref{H},\ref{barH}
and $M$ is the large scale $M>M_{X}$.  The idea is that when $H,
\bar{H}$ develop their VEVs such operators will become effective
Yukawa couplings of the form $h F \bar{F}$ with a small coefficient of
order $M_X^2/M^2$.  We shall only consider up to $n=2$ operators here,
since as we shall see even at this level there are a wealth of
possible operators that are encountered. Although we assume no
intermediate symmetry breaking scale (i.e. SU(4)$\otimes $SU(2)$_L
\otimes $SU(2)$_R$ is broken directly to the standard model at the
scale $M_X$) we shall allow the possibility that there are different
higher scales $M$ which are relevant in determining the operators.
For example one particular contraction of the indices of the fields
may be associated with one scale $M$, and a different contraction may
be associated with a different scale $M'$.  We shall either appeal to
this kind of idea in order to account for the various hierarchies
present in the Yukawa matrices, or to higher dimensional operators
which are suppressed by a further factor of $M$.

\subsection{A Simple $n=1$ Example}

In the present model, although there are no adjoint representations,
there will in general be non-renormalisable operators which closely
resemble those in SO(10) involving adjoint fields.  The simplest such
operators have already been considered in ref.\cite{422} and
correspond to $n=1$ in Eq.\ref{op}, with the $(H
\bar{H})$ group
indices contracted together.

These operators are similar to those of \cite{Larry} but with
$H\bar{H}$ playing the r\^{o}le of the adjoint Higgs representations.
It is useful to define the following combinations of fields,
corresponding to the different $H\bar{H}$ transformation properties
under the gauge group in Eq.\ref{422},
\begin{eqnarray}
(H \bar{H})_A & = & (1,1,1)
\nonumber \\
(H \bar{H})_B & = & (1,1,3)
\nonumber \\
(H \bar{H})_C & = & (15,1,1) \label{big} \\ (H \bar{H})_D & = &
(15,1,3)
\nonumber
\end{eqnarray}
The explicit form of the operators is given in Appendix 1.  For the
adjoint combinations we may write $B \equiv B^xT^x_R$, $C \equiv
C^pT^p$, $D \equiv D^{xp}T^x_RT^p$, where $T^x_R$ $(x=1,\ldots ,3)$
are the generators of SU(2)$_R$, and $T^p$ $(p=1,\ldots ,15)$ are the
generators of SU(4).  It is clear that when the fields $H,\bar{H}$
develop the VEVs in Eq.\ref{HVEV} the composite fields in Eq.\ref{big}
acquire VEVs
\begin{eqnarray}
<H \bar{H}>_A & = & <\nu_H><\bar{\nu}_H> \nonumber \\ <H \bar{H}>_B &
= & <\nu_H><\bar{\nu}_H>T^{3R} \nonumber \\ <H \bar{H}>_C & = &
<\nu_H><\bar{\nu}_H>T^{B-L} \nonumber \\ <H \bar{H}>_D & = &
<\nu_H><\bar{\nu}_H>T^{B-L}T^{3R} \label{AVEV}
\end{eqnarray}
where
\begin{eqnarray}
T^{B-L} & = & \frac{1}{2\sqrt{6}}\mbox{diag}(1,1,1,-3) \nonumber \\
T^{3R} & = & \frac{1}{2}\mbox{diag}(1,-1). \label{gens}
\end{eqnarray}

Armed with the above results it is straightforward to construct the
operators of the form of Eq.\ref{op} explicitly, and hence deduce the
effect of each operator.  For example for $n=1$ the four operators
are, respectively,
\begin{equation}
O^{A,B,C,D}_{ij}\sim F_i\bar{F}_jh\frac{(H
\bar{H})_{A,B,C,D}}{M^2}+H.c.
\label{n1}
\end{equation}
where we have suppressed gauge group indices.  When the combinations
$A,B,C,D$ in Eq.\ref{big} acquire the VEVs in Eq.\ref{AVEV} the
generators $T^{B-L}$ and $T^{3R}$ in Eq.\ref{gens} then just count the
quantum numbers of the components of the fields $F,\bar{F}$, leading
to quark-lepton and isospin splittings, as shown explicitly below:
\begin{eqnarray}
O^{A}_{ij} & = & a_{ij} (Q_{i}\bar{U}_{j}h_2 + Q_{i}\bar{D}_{j}h_1 +
L_{i}\bar{E}_{j}h_1 + H.c.) \nonumber \\ O^{B}_{ij} & = & b_{ij}
(Q_{i}\bar{U}_{j}h_2 - Q_{i}\bar{D}_{j}h_1 - L_{i}\bar{E}_{j}h_1 +
H.c.) \nonumber \\ O^{C}_{ij} & = & c_{ij} (Q_{i}\bar{U}_{j}h_2 +
Q_{i}\bar{D}_{j}h_1 -3L_{i}\bar{E}_{j}h_1 + H.c.) \nonumber \\
O^{D}_{ij} & = & d_{ij} (Q_{i}\bar{U}_{j}h_2 - Q_{i}\bar{D}_{j}h_1 +
3L_{i}\bar{E}_{j}h_1 + H.c.)
\label{effective}
\end{eqnarray}
where the coefficients of the operators $a_{ij},b_{ij},c_{ij},d_{ij}$
are all of order $\frac{<\nu_H><\bar{\nu}_H>}{M^2}$.
\subsection{Parity \label{sec:parity}}
In ref.\cite{422} combinations of the operators in Eq.\ref{effective}
were used to reproduce the successful RRR and GJ textures.  However it
is clear that there is no real justification for assuming that the
Yukawa matrices are symmetric in this model.  To illustrate the point,
let us impose a left-right (parity) symmetry on the model of the kind
introduced earlier in order to ensure that $g_L=g_R$ above $M_X$.
Under the $Z_2$ parity we have,
\begin{eqnarray}
A^{\mu}_L & \leftrightarrow & A^{\mu}_R \nonumber \\ F^i &
\leftrightarrow & \bar{F}^i \nonumber \\ H & \leftrightarrow & H'
\nonumber \\
\bar{H} & \leftrightarrow & \bar{H}' \nonumber \\
h & \leftrightarrow & h \label{parity}
\end{eqnarray}
where the fields $A^{\mu}_{L,R}$ are the gauge fields of SU(2)$_{L,R}$
and $H', \bar{H}'$ are $(4,2,1)$ and $(\bar{4}, \bar{2}, 1)$ irrep.s
respectively.  It is clear that operators such as those in Eq.\ref{op}
will not then lead to symmetric Yukawa matrices since under parity as
in Eq.\ref{parity}, the $(H\bar{H})$ combination which develops the
VEV and leads to the effective Yukawa coupling is transformed to
$(H'\bar{H}')$ which cannot attain a VEV if electroweak symmetry is to
remain intact at $M_X$ and so does not lead to an effective Yukawa
coupling. Note that this argument only applies to the
non-renormalisable operators. The renormalisable operators would lead
to symmetric Yukawa matrices if parity was imposed, since $h$ is
transformed into itself.  It is possible that there may be some
non-renormalisable operators which would lead to symmetric Yukawa
matrices at $M_{X}$ but these are not the kind of operators we
consider here. It is clear that the nature of this model is to lead to
non-symmetric Yukawa matrices, so the analysis of ref.\cite{422} must
be extended.
\subsection{General Analysis of $n=1$ and $n=2$ Operators}
The $n=1$ operators are by definition all of those operators which can
be constructed from the five fields $F \bar{F} h H \bar{H}$ by
contracting the group indices in all possible ways, as discussed in
Appendix 1. Here we only summarise the results of this analysis by
listing the group theoretical contractions of fields in
Table~\ref{table:opclass}, and the precise group structure of the
operators in Table~\ref{table:opcombs}. After the Higgs fields $H$ and
$\bar{H}$ develop VEVs at $M_X$ of the form $\langle H^{\alpha b}
\rangle = \langle H^{41}
\rangle = \nu_H$, $\langle \bar{H}_{\alpha x} \rangle = \langle
\bar{H}_{41} \rangle = \bar{\nu}_H$, the operators listed in the
appendix yield effective low energy Yukawa couplings with small
coefficients of order $M_X^2/M^2$. However, as in the simple example
discussed previously, there will be precise Clebsch relations between
the coefficients of the various quark and lepton component fields.
These Clebsch relations are summarised in Table~\ref{table:clebschs}.
Having discussed the origin of the effective Yukawa terms in some
detail for the $(H \bar{H})$ contracted operators, we shall now be
more schematic in our description of the remaining types of operator.

\begin{table}
\begin{center}
\def\arraystretch{1.4}
\begin{tabular}{|c|ccc|}
\hline
Operators & \multicolumn{3}{c|}{Combination}\\
\hline
$O^A$ to $O^D$ & $(F \bar{F})$ & $(H \bar{H})$ & $h$ \\ & $(15 \oplus
 1,2,2)$ & $(15 \oplus 1,1,1 \oplus 3)$ & $(1,2,2)$ \\
\hline
$O^E$ to $O^H$ & $(F H)$ & $(\bar{F} \bar{H})$ & $h$ \\ & $(6 \oplus
 10, 2, 2)$ & $(\bar{6} \oplus \bar{10}, 1, 1 \oplus
\bar{3})$ & $(1,\bar{2},2)$ \\
\hline
$O^I$ to $O^L$ & $(F \bar{H})$ & $(\bar{F} H)$ & $h$ \\ & $(15 \oplus
 1,2,\bar{2})$ & $(15 \oplus 1,1,1 \oplus 3)$ & $(1,\bar{2},2)$ \\
\hline
$O^K$ to $O^P$ & \multicolumn{3}{c|}{Mixed group structure} \\
\hline
\end{tabular}
\end{center}
\caption{Operator classification of the $n=1$ operators, including
those in Eq.\protect\ref{op},\protect\ref{big} (see Appendix 1 for
more details).}
\label{table:opclass}
\end{table}

\begin{table}
\begin{center}
\begin{tabular}{|c|c|c|c|}
\hline
Operator & SU(4)$_c$ & SU(2)$_L$ & SU(2)$_R$ \\
\hline
$O^A$ & $1 \otimes 1$ & $2 \otimes \bar{2}$ & $\bar{2} \otimes 1
\otimes 2$ \\

$O^B$ & $1 \otimes 1$ & $2 \otimes \bar{2}$ & $\bar{2} \otimes 3
\otimes 2$ \\

$O^C$ & $15 \otimes 15$ & $2 \otimes \bar{2}$ & $\bar{2} \otimes 1
\otimes 2$ \\

$O^D$ & $15 \otimes 15$ & $2 \otimes \bar{2}$ & $\bar{2} \otimes 3
\otimes 2$ \\

$O^E$ & $6 \otimes \bar{6}$ & $2 \otimes \bar{2}$ & $2
\otimes 2 \otimes 1$ \\

$O^F $&$ 6 \otimes \bar{6} $&$ 2 \otimes \bar{2}$&$ 2 \otimes 2
\otimes \bar{3} $  \\

$O^G $&$ 10 \otimes \bar{10}$&$ 2 \otimes \bar{2}$&$ 2 \otimes
\bar{2}$ \\

$O^H $&$ 10 \otimes \bar{10}$&$ 2 \otimes \bar{2}$&$ 2 \otimes 2
\otimes \bar{3}$  \\
$O^I $&$ 1 \otimes 1$&$ 2 \otimes \bar{2}$&$ \bar{2} \otimes 2$ \\
 $O^J $&$ 1 \otimes 1$&$ 2 \otimes \bar{2}$&$ 2 \otimes \bar{2}
 \otimes 3$ \\ $O^K $&$ 15 \otimes 15$&$ 2 \otimes \bar{2}$&$ 2
 \otimes \bar{2}$
\\
$O^L $&$ 15 \otimes 15$&$ 2 \otimes \bar{2}$&$ 2 \otimes \bar{2}
\otimes 3$. \\
\hline
\end{tabular}
\end{center}
\caption{The combinations shown in Table \protect\ref{table:opclass}
lead to the group structure for the $n=1$ operators shown here, where
the group singlet contraction is taken.}
\label{table:opcombs}
\end{table}

\begin{table}
\begin{center}
\begin{tabular}{|c|c|c|c|} \hline
 & $Q \bar{U} h_2$ & $Q \bar{D} h_1$ & $L \bar{E} h_1$ \\ \hline $O^A$
&1 & 1 & 1 \\ $O^B$ &1 & -1& -1 \\ $O^C$ &1 & 1 & -3 \\ $O^D$ &1 & -1&
3 \\ $O^E$ &0 & 1 & 0 \\ $O^F$ &1 & -4 & 0 \\ $O^G$ &0 & 1 & 2 \\
$O^H$ &2 & 1 & 2 \\ $O^I$ &0 & 0 & 0 \\ $O^J$ &0 & 0 & 1 \\
$O^{K,N,O}$ &1 & 0 &0 \\ $O^L$ &5 & 1 & 3/4 \\ $O^M$ &0 & 1 & 1 \\
$O^P$ &4 & 4 & 3 \\ \hline
\end{tabular}
\end{center}
\caption{When the Higgs fields develop their VEVs at $M_X$, the
$n=1$ operators lead to the effective Yukawa couplings with Clebsch
coefficients as shown. These results are a generalisation of
Eq.\protect\ref{effective}.}
\label{table:clebschs}
\end{table}
In Table~\ref{table:clebschs} we have neglected terms involving the
right handed neutrinos since they receive a large mass through the
see-saw mechanism\footnote{Which explains why $O^I \sim 0$, it only
contributes to a $\nu_R$ term. In this paper we shall not consider the
problem of neutrino masses.}. Note that associated with each operator
is only one coupling constant, so that for example $O^A$ gives the
same Yukawa coupling to the up and down quarks and the charged leptons
at $M_{X}$, as in Eq.\ref{effective}.
\begin{table}
\begin{center}
\begin{tabular}{|c|c|c|c|} \hline
 & $Q \bar{U} H_2$ & $Q \bar{D} H_1$ & $L \bar{E} H_1$ \\ \hline
$O^{Ad}$ &1 & 3& 9/4 \\ $O^{Dd}$ &1 & 3 & 3 \\ $O^{Md}$ &1 & 3& 6 \\
$O^1$ &0 & 1 & 1 \\ $O^2$ &0 & 1 & 3/4 \\ $O^3$ &0 & 1 & 2 \\ \hline
\end{tabular}
\end{center}
\caption{$n=2$ operators utilised, where $O^1$ is any one of
$O^{Dh,Dp,Dq,Dr,Ds}$; $O^2$ is any one of $O^{Ah,Ap,Aq,Ar,As}$ and
$O^3$ is one of $O^{Mh,Mp,Mq,Ms}$.  The operators are explicitly
defined in Appendix 2.}
\label{table:n2}
\end{table}
No non--renormalisable terms with $HH$ or $\bar{H} \bar{H}$ in the
Higgs structure are both gauge invariant and give a non zero mass
term, so only $(H \bar{H})$ operators are considered.

The $n=2$ operators are by definition all those operators which can be
constructed from $F \bar{F} h \bar{H} H \bar{H} H$ by contracting the
group indices in all possible ways, as discussed in Appendix 2.  There
are 400 $n=2$ operators, formed by different combinations of the
SU(4)$_c$ and SU(2)$_R$ structures e.g.\ we label an operator with
structures $A$ and $t$ as $O^{At}$. Brevity prevents us from listing
all the possible operators, but the useful operators are listed in
Table~\ref{table:n2}. Two features of the $n=2$ Clebsch coefficients
listed are useful: the operators $O^{Ad,Dd,Md}$ give the down quark
Clebsch coefficient to be three times that of the up quark, which
helps to predict a small $|V_{ub}|$, and the operators labeled
$O^{1,2,3}$ give masses to down quarks and leptons but not to up
quarks, helping to account for the up-down mass splitting in the first
family (see section 4.5).

\section{The Calculation}
\subsection{Masses and Mixing Angles at Low Energies}
To constrain the Yukawa matrices at $M_X$, we need to use
renormalisation group equations to evolve low energy parameters such
as CKM matrix elements and fermion masses up to $M_X$. We denote
running fermion masses at the $\overline{MS}$ scale $\mu$ as
$\bar{m}_f(\mu)$.
\begin{figure}
\begin{center}
\leavevmode
\hbox{\epsfxsize=4.5in
\epsfysize=2.6in
\epsffile{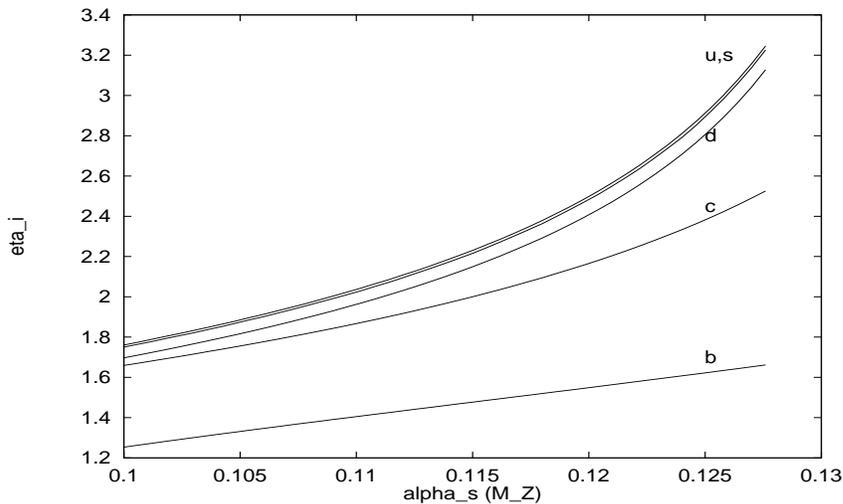}}
\end{center}
\caption{Running of Masses Between $\bar{m}_f$ and $\bar{m}_t$ (which
is displayed in Fig.\protect\ref{mtop}.)}
\label{etas}
\end{figure}

The masses of the fermions are first run up to the top mass
$\bar{m}_t$ using effective 3 loop QCD $\otimes$ 1 loop QED in the
$\overline{MS}$ renormalisation scheme
\cite{3loopQCDi,3loopQCDii,3loopQCDiii}.
Fig.\ref{etas} shows the parameter $\eta_i$ defined by
\begin{equation}
\eta_{f} =\frac{\bar{m}_{f}( \mbox{max}(\bar{m}_{f},1
\mbox{GeV}))}{\bar{m}_{f} ( \bar{m}_{t} ) },
\end{equation}
where max($\bar{m}_f$, 1 GeV) is the greater of $\bar{m}_f$ and 1 GeV.

For given values of $\tan \beta$, $\alpha_S (M_Z)$ and $\overline{MS}$
fermion masses defined by $\bar{m}_f \equiv \bar{m}_f($max$(\bar{m}_f,
1$GeV$))$, the diagonal Yukawa couplings at $\bar{m}_t$ are determined
by
\begin{eqnarray}
\lambda_{u,c,t} \left( \bar{m}_t \right) & = & \frac{\sqrt{2}
\bar{m}_{u,c,t}
}{v \eta_{u,c,t} \sin\beta}\label{Yuki}\\
\lambda_{d,s,b} \left( \bar{m}_t \right) & = & \frac{\sqrt{2}
\bar{m}_{d,s,b}}{v \eta_{d,s,b}  \cos \beta }\\
\lambda_{e,\mu,\tau} \left( \bar{m}_t \right) & = & \frac{\sqrt{2}
\bar{m}_{e,\mu,\tau}}
{v \eta_{e,\mu,\tau} \cos\beta}, \label{Yukii}
\end{eqnarray}
where $v=v_1^2 + v_2^2=246$ GeV.  All values used for the masses are
running values, as in ref.
\cite{RRR}\footnote{We would like to point out that a more sensible
convention to take would be to extract $\bar{m}_i(\sim~10$GeV$)$. This
could avoid threshold ambiguities when running up to $\bar{m}_t$ and
recent studies suggest that renormalons introduce an intrinsic
ambiguity into what one means by the pole mass of the lighter quarks
at these scales anyway.}.
\begin{table}
\begin{center}
\begin{tabular}{|c|c|c|} \hline
 & Lower Bound/GeV & Upper Bound/GeV \\ \hline $\bar{m}_d$ (1 GeV) &
0.0055 & 0.0115 \\ $\bar{m}_s$ (1 GeV)& 0.105 & 0.230 \\ $\bar{m}_u$
(1 GeV)& 0.0031 & 0.0064 \\ $\bar{m}_c(\bar{m}_c)$& 1.22 & 1.32 \\
$\bar{m}_b(\bar{m}_b)$ & 4.1 & 4.4 \\ \hline
\end{tabular}
\end{center}
\caption{Running masses of the lightest five quarks as provided by
\protect\cite{databook}.}
\label{table:masses}
\end{table}
The values of CKM mixing elements and running masses (of
Table~\ref{table:masses}) used are obtained from ref.
\cite{databook}:
\begin{equation}
V_{CKM} = \left(\begin{array}{ccc} 0.9747-0.9759 & 0.218-0.224 &
0.002-0.005 \\ 0.218-0.224 &0.9735-0.9751 &0.032-0.054\\ 0.003-0.018 &
0.030-0.054& 0.9985-0.9995\\
\end{array}\right).
\label{CKMmatrix}
\end{equation}
\begin{figure}
\begin{center}
\leavevmode
\hbox{\epsfxsize=4.5in
\epsfysize=2.6in
\epsffile{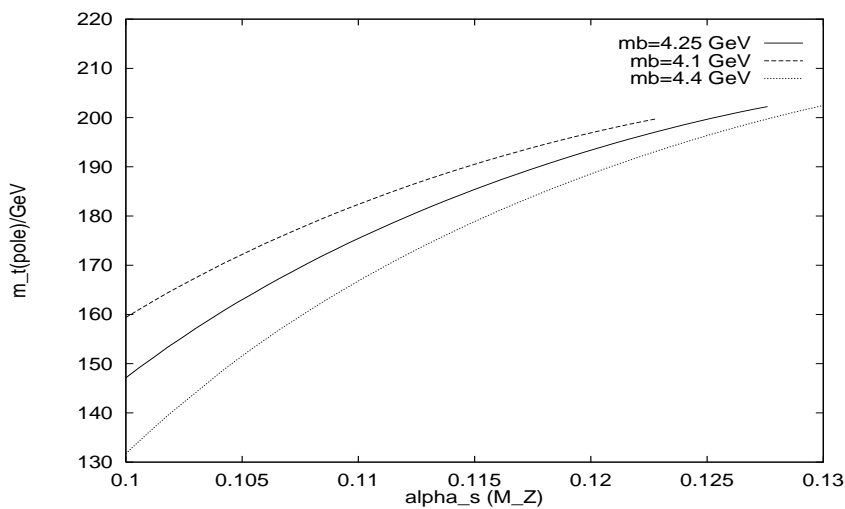}}
\end{center}
\caption{Physical mass on the top quark predicted by the
renormalisable 33 Yukawa term in Eq.\protect\ref{renorm} for various
values of $\bar{m}_b$. The pole mass is given by $m_t (\mbox{pole}) =
\bar{m}_t [ 1 + 4 \alpha_s ( \bar{m}_t)/(3 \pi) ]$ to one loop.}
\label{mtop}
\end{figure}
Lepton masses of course have no dependence on $\alpha_S$ to one loop,
and their $\eta$ values are tabulated in Table~\ref{table:leptons}.
\begin{table}
\begin{center}
\begin{tabular}{|c|c|c|c|}\hline
$\bar{m}_t$ / GeV & $\eta_e$ & $\eta_\mu$ & $\eta_\tau$ \\ \hline 140
& 1.018 & 1.018 & 1.016 \\ \hline 170 & 1.019 & 1.019 & 1.017 \\
\hline 200 & 1.020 & 1.020 & 1.018 \\ \hline
\end{tabular}
\end{center}
\caption{$\eta_i$ of the Leptons at various $\bar{m}_t$.}
\label{table:leptons}
\end{table}
\subsection{The Third Family: Yukawa Unification\protect\footnote{This
subject has been widely considered in the literature (cf.\
refs.\protect\cite{btauold,btaunew,Yuk}). We discuss it here for
completeness.}}

The third family have the largest and only renormalisable Yukawa term
in this scheme, which looks like
\begin{equation}
O_{33} = \lambda_{33} F^{\alpha a} \bar{F}_{\alpha x} h^x_a,
\label{renorm}
\end{equation}
where $\lambda_{33}$ is the universal Yukawa coupling at $M_{X}$,
unifying
\begin{equation}
\lambda_t (M_{X})= \lambda_b (M_{X}) = \lambda_\tau (M_{X}) =
\lambda_{33}.
\label{x3yukeq}
\end{equation}
$M_{X}$ is taken to be $10^{16}$ GeV. In fact, the results turn out to
be insensitive to whether we choose $M_{X} = 10^{16}$ or $10^{17}$
GeV, for example.

\begin{figure}
\begin{center}
\leavevmode
\hbox{\epsfxsize=4.5in
\epsfysize=2.6in
\epsffile{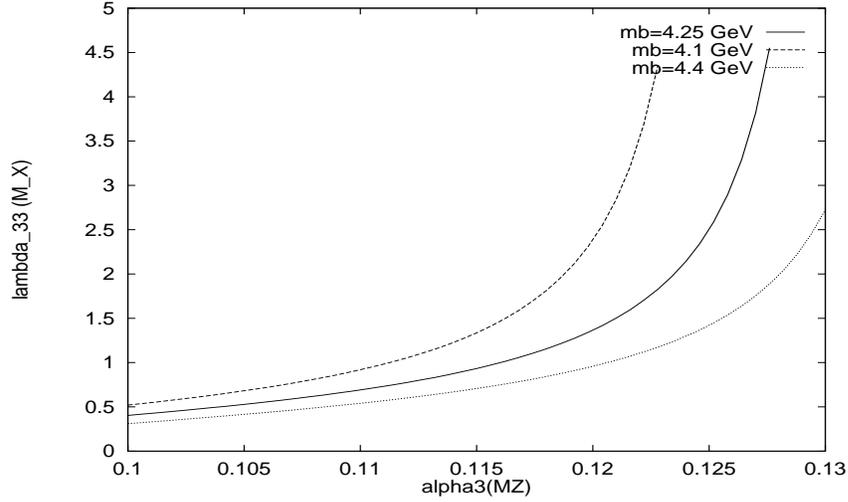}}
\end{center}
\caption{Renormalisable 3rd family Yukawa coupling $\lambda_{33}(M_X)$
for various values of $\bar{m}_b$}
\label{hgut}
\end{figure}

The third family Yukawa couplings are run from $\bar{m}_t$ to $M_{X}$.
This is achieved with the following SUSY one loop renormalisation
group (RG) equations (cf.\ ref.\cite{btaunew}):
\begin{eqnarray}
16 \pi^{2} \frac{\partial \lambda_{t}}{\partial t} &=& \lambda_{t}
\left[ 6\lambda_{t}^{2} +  \lambda_{b}^{2}  -
\left( \frac{13}{15}
g_{1}^{2} + 3g_{2}^{2} + \frac{16}{3}g_{3}^{2} \right)
\right] \nonumber \\
16 \pi^{2} \frac{\partial \lambda_{b}}{\partial t} &=&
\lambda_{b} \left[ 6
\lambda_{b}^{2} + \lambda_{\tau}^{2} + \lambda_{t}^{2} -
\left(
\frac{7}{15} g_{1}^{2} + 3g_{2}^{2} +
\frac{16}{3} g_{3}^{2} \right) \right] \nonumber \\
16 \pi^{2} \frac{\partial \lambda_{\tau}}{\partial t} &=&
\lambda_{\tau}
\left[ 4 \lambda_{\tau}^{2} + 3 \lambda_{b}^{2}
- \left( \frac{9}{5} g_{1}^{2} + 3g_{2}^{2} \right) \right],
\label{krg}
\end{eqnarray}
which are valid in the MSSM, assumed to be the correct theory between
$\bar{m}_t$ and $M_{X}$. We ignore low energy threshold corrections,
which should be smaller than the other theoretical uncertainties
involved (these are briefly presented in section 5.)

\begin{figure}
\begin{center}
\leavevmode
\hbox{\epsfxsize=4.5in
\epsfysize=2.6in
\epsffile{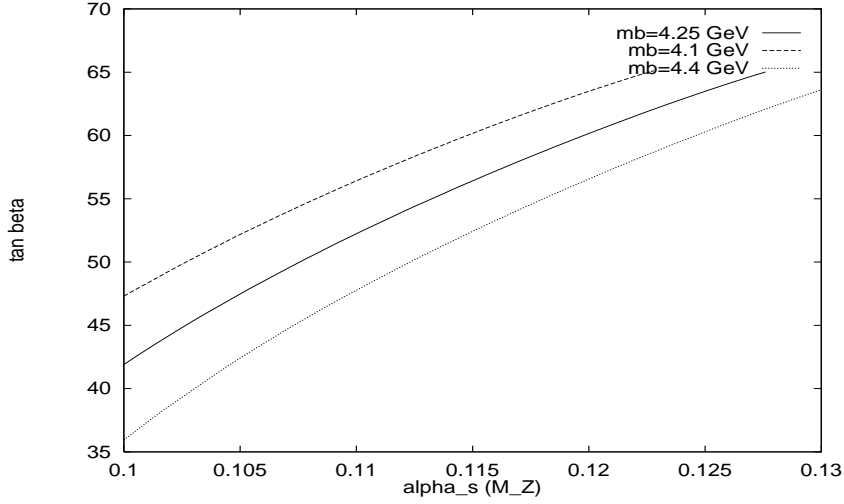}}
\end{center}
\caption{$\tan \beta$ prediction for various values of $\bar{m}_b$.}
\label{tanb}
\end{figure}

The procedure we follow for the third family is very similar to
ref.\cite{btaunew} for bottom-tau Yukawa unification, but are here
extended to include full top-bottom-tau Yukawa unification, according
to Eq.\ref{x3yukeq}. Briefly, the idea is to input $\bar{m}_b$,
$\bar{m}_\tau$ and $\alpha_S(M_Z)$ and then to predict $\bar{m}_t$ and
$\tan \beta$ as outputs using the constraint of Yukawa unification, as
in Eq.\ref{x3yukeq} after running the 3rd family Yukawa couplings up
to $M_X$. In practice this is complicated by the fact that the Yukawa
couplings $\lambda_t ( \bar{m}_t)$, $\lambda_b (\bar{m}_t ) $,
$\lambda_\tau (
\bar{m}_t ) $ all depend upon the input value of $\bar{m}_t$ and $\tan
\beta$. Thus we pick reasonable estimates of these quantities to input
into the RG routine. The output values obtained from this are fed back
as inputs until the iteration converges. In this way we select
$\bar{m}_t$, $\tan \beta$ consistent for Yukawa unification consistent
with given values of $\bar{m}_b$, $\bar{m}_\tau$,
$\alpha_S(M_Z)$. This results in the predictions for $m_t$(pole),
$\lambda_{33} (M_X)$ and $\tan \beta$ shown in
Figs.~\ref{mtop}-\ref{tanb}.

As Figs.~\ref{mtop}-\ref{tanb} illustrate, the results are highly
dependent upon $\alpha_s (M_Z)$, and fairly dependent on
$\bar{m}_b$. The value of $\tan
\beta$ is high
(35 to about 65) and $m_t($pole$)$ ranges from 130 to 190 GeV\footnote
{This is consistent with the CDF measurement of $m_t$ in
ref.\protect\cite{CDFtop}.}. Where the curves on the graphs stop for
high $\alpha_s (M_Z)$, one of the couplings has become too high and so
the model is not valid in this region of parameter space.  Note that
for certain superparticle spectrums, these results can be perturbed
because the determination of $\bar{m}_b(\bar{m}_b)$ does not include
certain one loop Yukawa corrections \cite{Larry}. This can have the
effect of lowering the prediction of $m_t$ by about 30 GeV, which is
still compatible with the CDF result for higher $\alpha_s$.
\subsection{First and Second family: Diagonal Yukawa Couplings}

In dealing with the first and second families we have to confront the
problem that the Yukawa matrices are not diagonal. As discussed widely
elsewhere \cite{Larry,RRR}, it is most convenient to diagonalise the
Yukawa matrices at $M_X$ before running them down to $\bar{m}_t$. It
is then possible to obtain RG equations for both the {\em diagonal}
\/Yukawa couplings $\lambda_{u,c,t}$, $\lambda_{d,s,b}$,
$\lambda_{e,\mu,\tau}$ and the Cabibbo-Kobayashi-Maskawa (CKM) matrix
elements $|V_{ij}|$\footnote{The empirical values of $|V_{ij}|$ were
taken to be at $\bar{m}_t$ instead of $M_Z$, introducing an error
whose magnitude is always less than 1 percent for our analysis.}
(ref.\cite{btaunew}). At one-loop these RG equations can be
numerically integrated so that the low energy physical couplings have
a simple scaling behaviour \cite{Larry}:
\begin{figure}
\begin{center}
\leavevmode
\hbox{\epsfxsize=4.5in
\epsfysize=2.6in
\epsffile{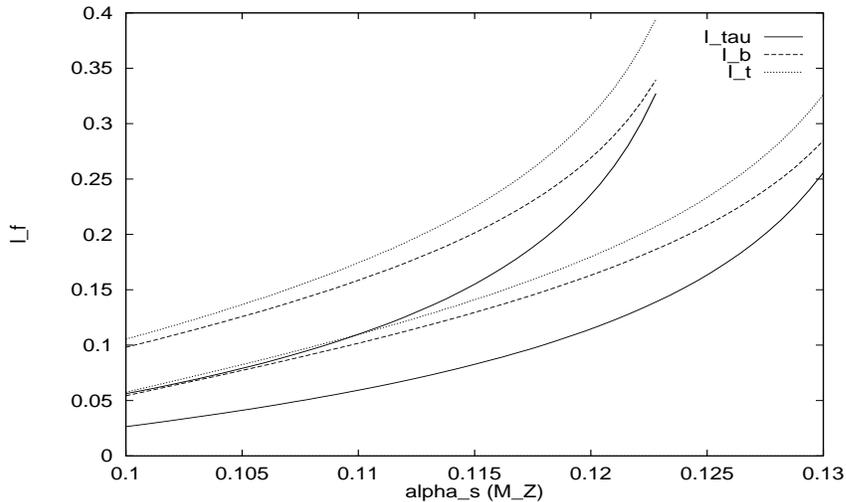}}
\end{center}
\caption{$I_f$ for the third family, as defined in
eq.\protect\ref{iint}. The lower (upper) triplet of curves is
associated with $\bar{m}_b$ = 4.1 (4.4) GeV. }
\label{Is}
\end{figure}

\begin{eqnarray}
\left( \frac{\lambda_{u,c}}{\lambda_{t}} \right)_{\bar{m}_t} & = &
\left( \frac{\lambda_{u,c}}
{\lambda_{t}} \right)_{M_{X}} e^{3I_{t} + I_{b}} \label{start}
\nonumber \\
\left( \frac{\lambda_{d,s}}{\lambda_{b}} \right)_{\bar{m}_t} & = &
\left( \frac{\lambda_{d,s}}
{\lambda_{b}} \right)_{M_{X}} e^{3I_{b} + I_{t}} \nonumber \\
\left( \frac{\lambda_{e, \mu}}{\lambda_{\tau}} \right)_{\bar{m}_t} & =
& \left(
\frac{\lambda_{e, \mu}}
{\lambda_{\tau}} \right)_{M_{X}} e^{3I_{\tau}} \nonumber \\
\frac{ \mid V_{cb} \mid _{M_{X}}}{\mid V_{cb} \mid
_{\bar{m}_t} } & = & e^{I_{b}+I_{t}} , \label{gutsusy}
\end{eqnarray}
with identical scaling behaviour to $V_{cb}$ of $V_{ub}$, $V_{ts}$,
$V_{td}$, where
\begin{equation}
I_{i}=\int _{\ln \bar{m}_t}^{\ln {M_{X}}} \left( \frac{\lambda_{i}
\left( t
\right) }{4 \pi} \right)^2 dt, \label{iint}
\end{equation}
and $t=\ln \mu$, $\mu$ being the $\overline{MS}$ scale.  To a
 consistent level of approximation $V_{us}$, $V_{ud}$, $V_{cs}$,
 $V_{cd}$, $V_{tb}$, $\lambda_{u}$/$\lambda_{c}$,
 $\lambda_{d}$/$\lambda_{s}$ and $\lambda_{e}$/$\lambda_{\mu}$ are RG
 invariant.  The CP violating quantity J scales as $V_{cb}^{2}$. The
 $I$ integrals of the third family are displayed in Fig.\ref{Is} for
 the allowed range of $\bar{m}_b$ and $\alpha_S (M_Z)$.
\begin{figure}
\begin{center}
\leavevmode
\hbox{\epsfxsize=4.5in
\epsfysize=2.6in
\epsffile{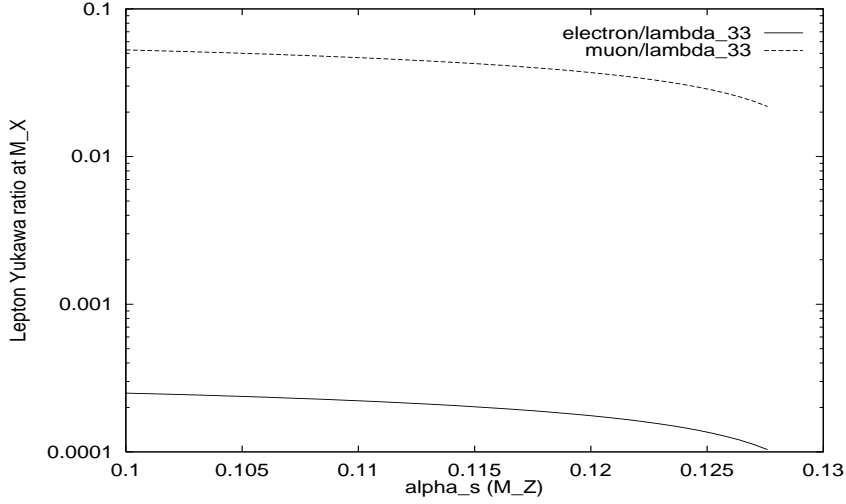}}
\end{center}
\caption{Empirically derived ratios of first and second to third family
lepton Yukawa couplings at $M_X$ for $\bar{m}_b=4.25$ GeV.}
\label{fammag}
\end{figure}

\begin{figure}
\begin{center}
\leavevmode
\hbox{\epsfxsize=4.5in
\epsfysize=2.6in
\epsffile{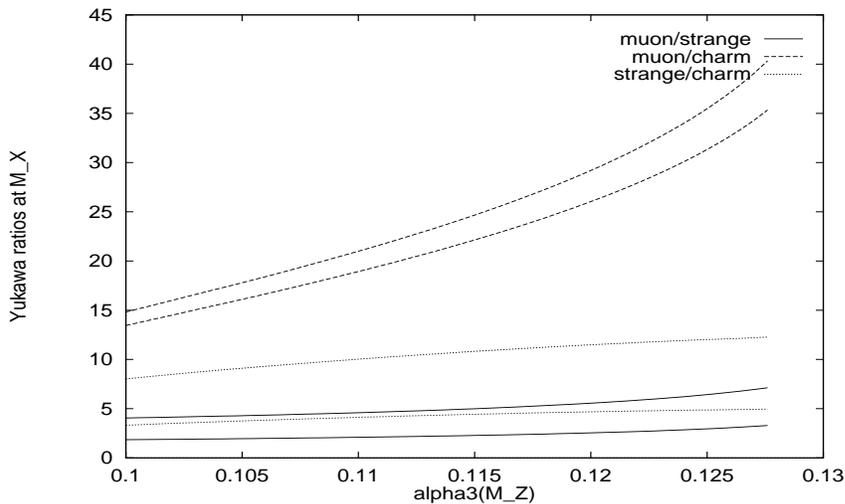}}
\end{center}
\caption{Empirical bounds on ratios of diagonal Yukawa couplings of
the second family at $M_{X}$ and $\bar{m}_b=$ 4.25 GeV.}
\label{heavy}
\end{figure}

\begin{figure}
\begin{center}
\leavevmode
\hbox{\epsfxsize=4.5in
\epsfysize=2.6in
\epsffile{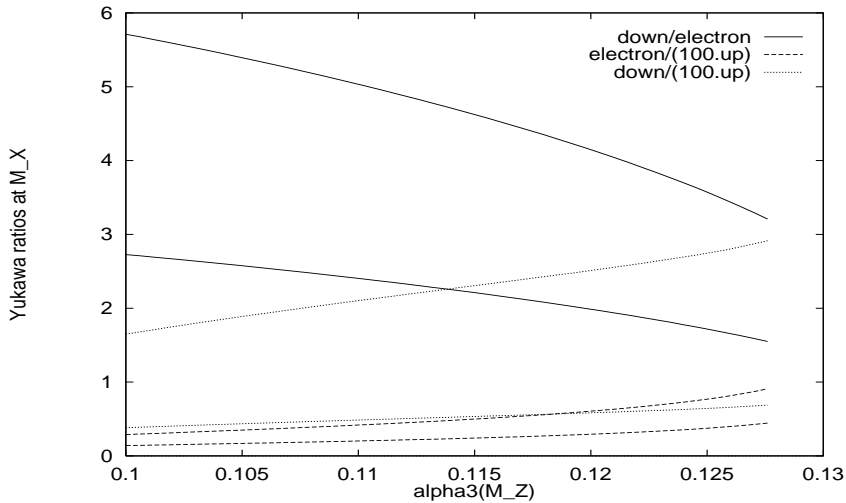}}
\end{center}
\caption{Empirical bounds on ratios of diagonal Yukawa
couplings of the first family at $M_{X}$ using $\bar{m}_b=4.25$ GeV.}
\label{light}
\end{figure}

\begin{figure}
\begin{center}
\leavevmode
\hbox{\epsfxsize=4.5in
\epsfysize=2.6in
\epsffile{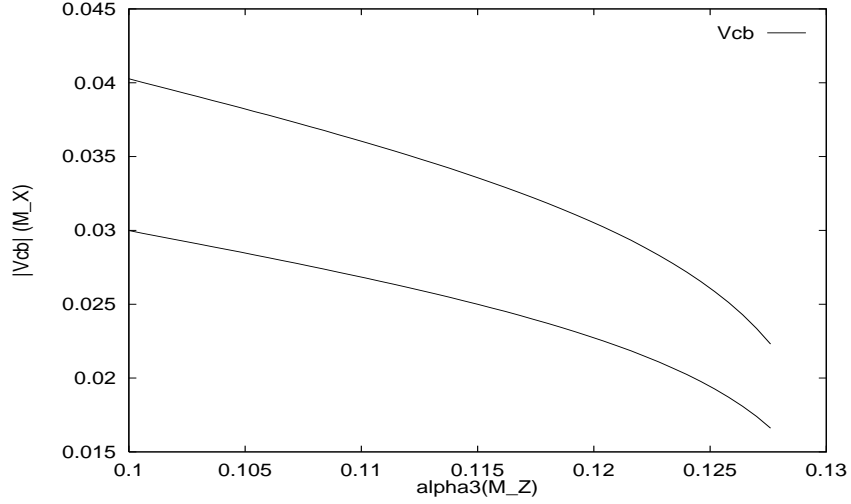}}
\end{center}
\caption{$|V_{cb}|$ calculated at $M_{X}$ from the low energy value.}
\label{Vcb}
\end{figure}

Using Eqs.\ref{Yuki}-\ref{Yukii} we may determine the diagonal Yukawa
couplings in this model at $\bar{m}_t$. The third family Yukawa
couplings at $M_X$ (all equal) are given by Fig.\ref{hgut}. The first
and second family diagonal Yukawa couplings at $M_X$ are then easily
obtained from the scaling relations in Eq.\ref{gutsusy}, using the
$I_f$ integrals in Fig.\ref{Is}. These GUT scale Yukawa couplings
expressed as ratios are shown in
Figs.~\ref{fammag},\ref{heavy},\ref{light}. The relative magnitude of
the diagonal Yukawa couplings between the first two families and the
third is displayed in Fig.\ref{fammag}, where the ratio should be
$O(\epsilon)$ for the first family and $O(\epsilon^2)$ for the second,
if the assumption of suppression of the mass scales is to be
correct. As seen in the figure, this identifies $\log ( \epsilon) \sim
O(-1.5)$.
The second family couplings at $M_X$ are shown in Fig.\ref{heavy} and
show that $\lambda_s : \lambda_\mu : \lambda_c \sim 6 : 20 : 1$. The
first family couplings are displayed in Fig.\ref{light} and give
$\lambda_d : \lambda_e : \lambda_u \sim 3 : 1 : 1/30$. These are the
patterns that must be replicated by our model if it going to be
phenomenologically viable. In Fig.\ref{Vcb} we also show the absolute
value of $|V_{cb}|$ at $M_X$, calculated by running the value at $M_Z$
in Eq.\ref{CKMmatrix} up to $M_X$ using Eq.\ref{gutsusy}.
\subsection{The Lower 2 by 2 Block of The Yukawa Matrix: The
Prediction of the Strange Quark Mass} The heavy 2 by 2 sectors of the
Yukawa matrices are considered first, in isolation from the rest of
the Yukawa matrix. This is possible because we shall assume that the
Yukawa matrices at $M_{X}$ are all of the form
\begin{equation}
Y^{U,D,E} = \left(
\begin{array}{ccc}
O(\epsilon^2) & O(\epsilon^2) & 0 \\ O(\epsilon^2) & O(\epsilon) &
O(\epsilon) \\ 0 & O(\epsilon) & O(1) \\
\end{array}\right),
\label{matrixform}
\end{equation}
where $\epsilon << 1$ and some of the elements may have approximate or
exact texture zeroes in them. Assuming the form Eq.\ref{matrixform}
allows us to consider the lower 2 by 2 block of the Yukawa matrices
first.  In diagonalising the lower 2 by 2 block separately, we
introduce corrections of order $\epsilon^2$ and so the procedure is
consistent to first order in $\epsilon$.

In searching for the minimum number of operators that fit
phenomenological constraints (so that maximum predictivity is
attained) there are some general arguments that yield a lower bound on
the number of operators.  It is clear that there must be at least 3
operators in the lower 2 by 2 block since it should have no zero rows
or columns, or its determinant would be zero, implying that it
contains a zero eigenvalue and therefore a zero mass. Also, for a
non-zero $|V_{cb}|$, we require there to be an operator in the 32
position in the down and/or up matrices. In fact we shall require 4
operators in the lower 2 by 2 block since the Clebsch coefficients
listed in Table~\ref{table:clebschs} are not sufficient to describe
successfully all the features of Fig.\ref{heavy}, in particular the
charm-muon splitting ($\lambda_\mu$/$\lambda_c \sim 18)$\footnote{The
discussion of complex phases is left to later. Here we assume that all
of the Yukawa parameters are real, which happens to make no difference
to our analysis of the lower 2 by 2 block (see Section 4.4).}. With 3
operators, excluding the third family discussed previously, we have
three Yukawa coupling parameters, not including the 33 coupling which
is already calculated. There are four observables connected with these
parameters, $|V_{cb}|$, $m_s$, $m_c$ and $m_\mu$.  We use $|V_{cb}|,
\bar{m}_c, \bar{m}_\mu$ as inputs but the only prediction we can make
is for the strange mass $\bar{m}_s$.

Given the constraints listed above, the only ansatze that successfully
satisfy them are listed here:
\begin{eqnarray}
A_1 &=& \left[\begin{array}{cc} O^D - O^C & 0 \\ O^{B} & O_{33}
\end{array}\right] \label{A1} \\ A_2 &=& \left[\begin{array}{cc} 0 &
O^A - O^B \\ O^D & O_{33}
\end{array}\right]  \\
A_3 &=& \left[\begin{array}{cc} 0 & O^C - O^D \\ O^B & O_{33}
\end{array}\right]  \\
A_4 &=& \left[\begin{array}{cc} 0 & O^{C} \\ O^A - O^B & O_{33}
\end{array}\right]  \\
A_5&=& \left[\begin{array}{cc} 0 & O^{A} \\ O^C - O^D & O_{33}
\end{array}\right] \\
A_6&=& \left[\begin{array}{cc} O^{K} & O^{C} \\ O^M & O_{33}
\end{array}\right]\\
A_7&=& \left[\begin{array}{cc} O^{K} & O^G \\ O^G & O_{33}
\end{array}\right]  \\
A_8&=& \left[\begin{array}{cc} 0 & O^H \\ O^G - O^{K} & O_{33}
\end{array}\right].
\label{endans}
\end{eqnarray}
Note that $O^D$ can be used in the 32 position of $A_1$ instead of
$O^B$, yielding exactly the same results. Also, ansatze replacing
$O^C$ with $O^D$ in $A_4$, $O^A$ with $O^B$ in $A_5$, $O^K$ with
$O^{N}$ or $O^O$ in $A_{6-8}$ and $O^C$ with $O^D$ in $A_{6,7}$ also
yield the same results. $A_7$ may be transposed to yield the same
results, as may $A_6$ if $O^C$ is not in the 23 position before
transposing, as this would predict zero $|V_{ub}|$.

{}From the above ansatze in Eqs.\ref{A1}-\ref{endans}, the ratio of
muon to strange Yukawa couplings at $M_X$ is found to be:
\begin{equation}
\left( \frac{\lambda_\mu}{\lambda_s} \right)_{M_{X}} \equiv l.
\label{2ndpred}
\end{equation}
where $l$ is a ratio of Clebsch coefficients, predicted to be $l=3$
(as in the GJ ansatz) or $l=4$ (a new prediction).

Ansatze $A_{1-6}$ predict $l=3$ in Eq.\ref{2ndpred} and ansatze
$A_{7,8}$ predict $l=4$.

\begin{figure}
\begin{center}
\leavevmode
\hbox{\epsfxsize=4.5in
\epsfysize=2.6in
\epsffile{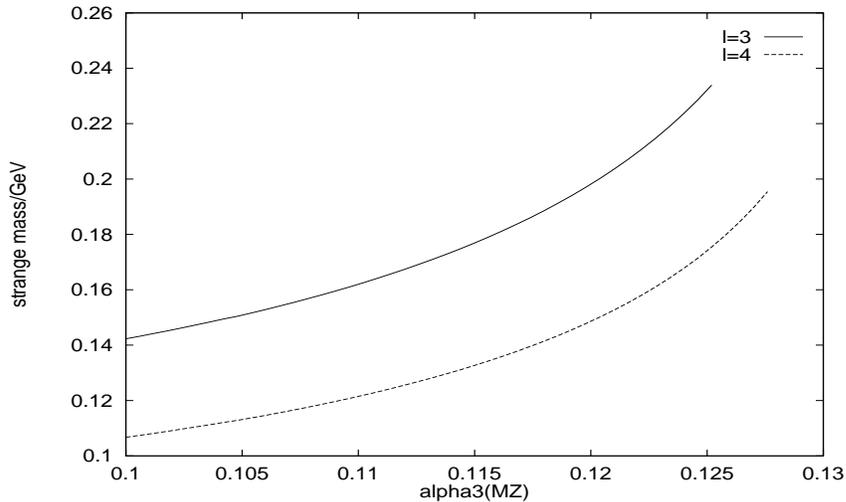}}
\end{center}
\caption{$\bar{m}_s$(1 GeV) predictions as a result of
Eq.\protect\ref{2ndpred}, to be compared with the experimental value
quoted in Table
\protect\ref{table:masses}.}
\label{ms}
\end{figure}

Eq.\ref{2ndpred} gives a prediction of the quantity $\lambda_s
(M_{X})$, which can in turn be run down to 1 GeV, yielding a
prediction for $\bar{m}_s$.  Fig.\ref{ms} shows the predictions for
$\bar{m}_s$(1 GeV), which were found to have negligible dependence
upon $\bar{m}_b$. Note that the $l=4$ curve in Fig.\ref{ms}
corresponding to ansatze $A_{7,8}$ works just as well as the $l=3$ GJ
prediction.

\subsection{The Upper 2 by 2 Block of The Yukawa Matrix: The
Prediction of the Down Quark Mass.}  Having diagonalised the lower 2
by 2 block, we have an effective Yukawa coupling in the 22 position
for each of the Yukawa matrices. We must now add additional operators
to account for $m_e, m_u, m_d, |V_{ub}|$ and $|V_{us}|$. We shall
assume that these additional operators are $n=2$ ones, as discussed
previously together with even smaller $n=3$ operators which we denote
generically $O^{n=3}$.  Naively the minimum number of additional
operators in the upper 2 by 2 block is 2: one in the 21 position to
account for $|V_{us}|$ and one on the top row to avoid a massless
first family. However, it is impossible to account for $(\lambda_u /
\lambda_e )_{M_X} \sim 1/30$ and $(\lambda_u /
\lambda_d)_{M_X} \sim 1/100$ using only two operators, since the
magnitude of the Clebsch coefficients in Table~\ref{table:clebschs}
are simply not big enough. To circumvent this problem we shall use the
operators in Table~\ref{table:clebschs} which give a zero contribution
to the up mass and a non zero contribution to the down and electron
mass, namely $O^{1,2,3}$. In order to achieve a non-zero $|V_{ub}|$ we
require an operator in the 21 position which gives a non-zero
contribution to the up-matrix. To provide a phenomenologically viable
$|V_{ub}|$ it turns out (see later) that the Clebsch of the down
Yukawa coupling in the 21 position has to be 3 times that of the up
Yukawa coupling in the 21 position. This implies that the 21 operator
must be one of $O^{Ad}, O^{Dd}$ or $O^{Md}$. If the 12 operator is
simply $O^{1,2,3}$ then this predicts a massless up quark. In order to
obtain a small up quark mass we must add a small third operator
(denoted $O^{n=3}$) to the 11 or 12 positions\footnote{In all of the
ansatze $B_{1-8}$, the $O^{n=3}$ operator can be placed in the 11
position yielding identical results.}. This leads to the successful
upper 2 by 2 block ansatze shown below.  With three extra operators in
the upper block we have three Yukawa coupling parameters, connected
with the 5 observables $|V_{us}|, m_u, m_d, m_e$ and
$|V_{ub}|$. $m_u$, $m_e$ and $|V_{us}|$ are used as inputs and $m_d,
|V_{ub}|$ are predicted\footnote{However, there appears an unremovable
phase that gives CP violation, so the prediction of $|V_{ub}|$ depends
upon this phase (this is explained more comprehensively in section
4.6).}.

The possible ansatze in the down sector that account for correct
$|V_{us}|, m_u, m_d, m_e$ and $|V_{ub}|$ and {\em also} \/generate CP
violation are:
\begin{eqnarray}
B_1 &=& \left[\begin{array}{cc} 0 & O^{n=3} + O^1\\ O^{Ad} & X
\end{array}\right]\label{suclighti}\\ B_2 &=& \left[\begin{array}{cc}
0 & O^{n=3} + O^2\\ O^{Ad} & X \end{array}\right]\\ B_3 &=&
\left[\begin{array}{cc} 0 & O^{n=3} + O^3\\ O^{Ad} & X
\end{array}\right]\\ B_4 &=& \left[\begin{array}{cc} 0 & O^{n=3} +
O^1\\ O^{Dd} & X \end{array}\right]\\ B_5 &=& \left[\begin{array}{cc}
0 & O^{n=3} + O^2\\ O^{Dd} & X \end{array}\right]\\ B_6 &=&
\left[\begin{array}{cc} 0 & O^{n=3} + O^3\\ O^{Dd} & X
\end{array}\right]\\ B_7 &=& \left[\begin{array}{cc} 0 & O^{n=3} +
O^1\\ O^{Md} & X \end{array}\right]\\ B_8 &=& \left[\begin{array}{cc}
0 & O^{n=3} + O^2\\ O^{Md} & X \end{array}\right]. \label{suclightii}
\end{eqnarray}
$X$ stands for whatever is left in the 22 position, after the lower 2
by 2 submatrix has been diagonalised.  Each of the successful ansatze
gives a prediction for the down Yukawa coupling at $M_X$ in terms of
the electron Yukawa coupling:
\begin{equation}
\left( \frac{\lambda_d}{\lambda_e} \right)_{M_{X}} \equiv k,
\label{1stpred}
\end{equation}
where $k=3$ is the Georgi--Jarlskog prediction of $\bar{m}_d$. Other
viable possibilities found in our analysis are
$k=2,4,\frac{8}{3},\frac{16}{3}$ as shown in Table~\ref{kvals}.  The
$k$ values as defined in Eq.\ref{1stpred} depend upon the $l$ value in
the heavy submatrix and are displayed in Table~\ref{kvals}.
\begin{table}
\begin{center}
\begin{tabular}{|c|c|c|c|c|c|c|c|c|} \hline
$k$ & $B_1$ & $B_2$ & $B_3$ & $B_4$ & $B_5$ & $B_6$ & $B_7$ & $B_8$ \\
\hline
$A_{1-6}$ & 4 & 16/3 & 2 & 3 & 4 & (3/2) & (3/2) & 2 \\ \hline
$A_{7,8}$ & 16/3 & (64/9) & 8/3 & 4 & 16/3 & 2 & 2 & 8/3 \\ \hline
\end{tabular}
\end{center}
\caption{$k$ values predicted by the ansatze $A_1$ to $A_8$ when
combined with $B_1$ to $B_8$. Note that the bracketed entries predict
$\bar{m}_d (1$ GeV$)$ to be outside the empirical range, and so are
not included in Fig.\protect\ref{md}.}
\label{kvals}
\end{table}

\begin{figure}
\begin{center}
\leavevmode
\hbox{\epsfxsize=4.5in
\epsfysize=2.6in
\epsffile{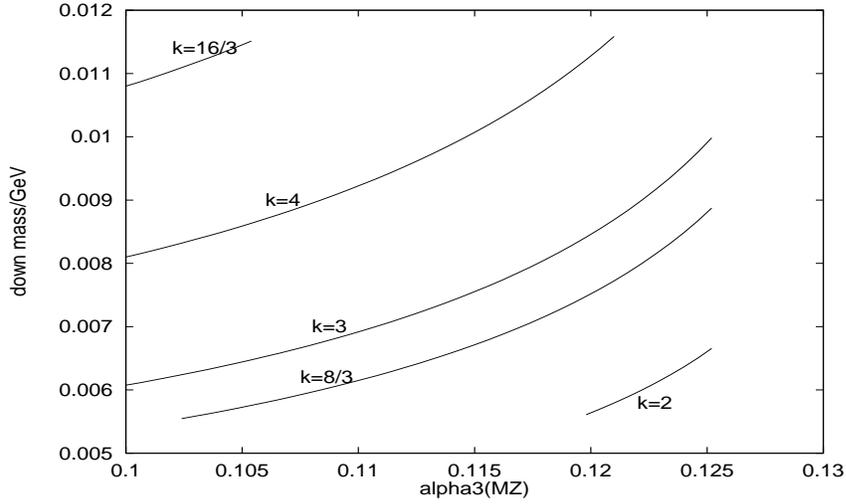}}
\end{center}
\caption{$\bar{m}_d$(1 GeV) prediction as a result of the
Eq.\protect\ref{1stpred}, with the $k$ values outlined in
Table~\protect\ref{kvals}, labeling each prediction.}
\label{md}
\end{figure}

Fig.\ref{md} shows the five possible predictions for $\bar{m}_d$ in
the separate ansatze. Again, the results were found to have negligible
dependence on $\bar{m}_b$ because this will only change the RG running
by affecting $\bar{m}_t$. The large $\tan \beta$ dependence factors
out since we relate the mass of the down quarks to the mass of the
leptons only, and these have the same dependence upon $\tan \beta$.
It should be made clear that up until now all of the Yukawa couplings
are assumed to be real and positive, but in order to do a full
analysis of the CKM matrix including the CP violating phase we shall
now drop this assumption.

\subsection{The Full 3 by 3 Yukawa Matrix: The Prediction of
$|V_{ub}|$ in Terms of the CP-Violating Phase} So far we have
discussed the lower 2 by 2 block and the upper 2 by 2 block of the
Yukawa matrices, assuming them to be of the form shown in
Eq.\ref{matrixform}, and taking all of the couplings to be real and
positive. In fact the effective Yukawa couplings must be regarded as
complex parameters with relative phases between them. This does not
affect our analysis of third family Yukawa unification. Nor does it
affect the predictions of the strange and down quark masses which
follow from Eqs.\ref{1stpred},\ref{2ndpred}, since $l$ and $k$ are
simply ratios of Clebsch coefficients. However the precise list of
successful lower 2 by 2 block ansatze $A_i$ in section 4.4, and the
upper 2 by 2 block ansatze $B_i$ in section 4.5 was in fact based on a
full analysis of the 3 by 3 Yukawa matrices, including complex phases.
We shall not repeat the full analysis which led to the successful
ansatze $A_i$ and $B_i$ here. However in order to illustrate our
approach and discuss the remaining CKM parameters, it is instructive
to consider one example as described below.

The successful ansatze consist of any of the lower 2 by 2 blocks $A_i$
combined with any of the upper 2 by 2 blocks $B_i$, subject to the
restrictions shown in Table~\ref{kvals}.  For example let us consider
$A_1$ in the lower 2 by 2 block combined with any of the $B_i$ in the
upper 2 by 2 block, focusing particularly on $A_1$ combined with
$B_1$. Just above $M_X$, before the $H,\bar{H}$ fields develop VEVs,
we have the operators
\begin{equation}
\left[\begin{array}{ccc}
0 & O^1_{12} + {O}_{12}^{n=3} & 0 \\ O^{Ad}_{21} & O^D_{22} -
{O}^C_{22} & 0 \\ 0 & O^B_{32} & O_{33} \\ \end{array}\right],
\end{equation}
which implies that at $M_X$ the Yukawa matrices are of the form
\begin{equation}
Y^{U,D,E} =
\left[
\begin{array}{ccc}
0 & H_{12}x_{12}^{U,D,E} e^{i \phi_{12}}+H_{12}{'} x{'}_{12}^{U,D,E}
e^{i \phi_{12}'}& 0 \\ H_{21}x_{21}^{U,D,E} e^{i \phi_{21}} &
H_{22}x_{22}^{U,D,E} e^{i
\phi_{22}} -
H_{22}' x_{22}'^{U,D,E} e^{i \phi_{22}'} & 0 \\ 0 & H_{32}
x_{32}^{U,D,E} e^{i \phi_{32}} & H_{33}e^{i \phi_{33}} \\
\end{array}\right],
\label{firstyuk}
\end{equation}
where we have factored out the phases of the operators and $H_{ik}$
are the magnitudes of the coupling constant associated with
$O_{ik}$. Note the real Clebsch coefficients $x_{ik}^{U,D,E}$ give the
splittings between $Y^{U,D,E}$. In our particular case $A_1, B_1$ they
are given by
\begin{eqnarray}
x_{12}^U =0 & x_{12}^D = 1 & x_{12}^E = 1 \nonumber \\ x_{21}^U =1 &
x_{21}^D = 3 & x_{21}^E = 9/4 \nonumber \\ x_{22}^U =1 & x_{22}^D = -1
& x_{22}^E = 3 \nonumber \\ {x'}_{22}^U = 1 & {x'}_{22}^D = 1 &
{x'}_{22}^E = -3 \nonumber \\ x_{32}^U = 1 & x_{32}^D = -1 & x_{32}^E
= -1 \label{clebsch2}
\end{eqnarray}
and ${x'}_{12}^U \neq 0$.  We now make the transformation in the 22
element of Eq.\ref{firstyuk}
\begin{eqnarray}
x_{22}^U H_{22} e^{i \phi_{22}} - x{'}_{22}^U H_{22}' e^{i \phi'_{22}}
&\equiv& H_{22}^U e^{i \phi_{22}^U} \nonumber \\ x_{22}^D H_{22} e^{i
\phi_{22}} - x{'}_{22}^D H_{22}' e^{i \phi'_{22}} &\equiv& H_{22}^D
e^{i \phi_{22}^D} \nonumber \\ x_{22}^E H_{22} e^{i \phi_{22}} -
x{'}_{22}^E H_{22}' e^{i \phi'_{22}} &\equiv& H_{22}^E e^{i
\phi_{22}^E}, \label{transform22}
\end{eqnarray}
where $H_{22}^{U,D,E}, \phi_{22}^{U,D,E}$ are real positive
parameters. It follows from the Clebsch structure in Eq.\ref{clebsch2}
that $H_{22}^E = 3 H_{22}^D$ and $\phi_{22}^E = \phi_{22}^D$. In
general we shall write $H_{22}^E = l H_{22}^D$, where $l=3$ in this
case.

At $M_X$, we have the freedom to rotate the phases of $F^i$ and
$\bar{F}_j$, since this leaves the lagrangian of the high energy
theory invariant. In doing this we rotate away 5 phases in the
matrices since there are only 5 {\em relative} \/phases:
\begin{eqnarray}
\left[\begin{array}{c}
\bar{F}_1 \\ \bar{F}_2 \\ \bar{F}_3 \end{array}\right] &\rightarrow&
\left[\begin{array}{ccc}
e^{-i (\phi_{32} - \phi_{12})} & 0 & 0 \\ 0 & e^{-i (\phi_{32} -
\phi_{22}^D)} & 0 \\ 0 & 0 & 1 \\ \end{array}\right]
\left[\begin{array}{c}
\bar{F}_1 \\ \bar{F}_2 \\ \bar{F}_3 \end{array}\right]
\nonumber \\
\left[\begin{array}{c}
F_1 \\ F_2 \\ F_3 \end{array}\right] &\rightarrow&
\left[\begin{array}{ccc}
e^{-i (-\phi_{32}+\phi_{22}^D - \phi_{21})} & 0 & 0 \\ 0 &e^{i
\phi_{32}} & 0 \\ 0 & 0 & e^{i \phi_{33}} \\ \end{array}\right]
\left[\begin{array}{c}
F_1 \\ F_2 \\ F_3 \end{array}\right].
\label{phaserots}
\end{eqnarray}

Below $M_X$, the multiplets $F, \bar{F}$ are no longer connected by
the gauge symmetry in the effective field theory, since it is The
Standard Model.  We now define our notation as regards the effective
field theory below $M_X$ as follows. The effective quark Yukawa terms
are written (suppressing all indices)
\begin{equation}
(U_R)^c Y^U Q_L h_2 + (D_R)^c Y^D Q_L h_1 + H.c.
\label{quarkL}
\end{equation}
We transform to the quark mass basis by introducing four 3 by 3
unitary matrices $V_{U_{L,R}}, V_{D_{L,R}}$ then the Yukawa terms
become
\begin{equation}
(U_R)^c V_{U_R}^\dagger V_{U_R} Y^U V_{U_L}^\dagger V_{U_L} Q_L h_2 +
(D_R)^c V_{D_R}^\dagger V_{D_R} Y^D V_{D_L}^\dagger V_{D_L} Q_L h_1 +
H.c.
\label{massbasis}
\end{equation}
where $Y^U_{\mbox{diag}} = V_{U_R} Y^U V_{U_L}^\dagger$ and
$Y^D_{\mbox{diag}} = V_{D_R} Y^D V_{D_L}^\dagger$ are the diagonalised
Yukawa matrices.  With the definitions in
Eqs.\ref{quarkL},\ref{massbasis}, the CKM matrix is of the form
\begin{equation}
V_{CKM} \equiv V_{U_L} V_{D_L}^\dagger. \label{CKM}
\end{equation}

In all of the cases considered, $x_{12}^U=0$ and $x_{12}'^{D,E}=0$ so
that the Yukawa matrices which result from
Eqs.\ref{firstyuk},\ref{transform22},\ref{phaserots} are
\begin{eqnarray}
Y^D &=& \left[\begin{array}{ccc} 0 & H_{12}x_{12}^D & 0 \\ H_{21}
x_{21}^D & H_{22}^D& 0 \\ 0 & H_{32}x_{32}^D & H_{33} \\
\end{array}\right] \nonumber \\
Y^E &=& \left[\begin{array}{ccc} 0 & H_{12}x_{12}^E & 0 \\ H_{21}
x_{21}^E & l H_{22}^D & 0 \\ 0 & H_{32}x_{32}^E & H_{33} \\
\end{array}\right] \nonumber \\
Y^U &=& \left[\begin{array}{ccc} 0 & H_{12}'{x'}^U_{12} e^{i
(\phi_{12}{'} - \phi_{12})} & 0 \\ x_{21}^U H_{21} & H_{22}^U e^{i
(\phi_{22}^U- \phi_{22}^D)} & 0 \\ 0 & H_{32} x_{32}^U & H_{33} \\
\end{array}\right].
\end{eqnarray}
In order to diagonalise the quark Yukawa matrices, we first make $Y^U$
real, by multiplying by phase matrices
\begin{eqnarray}
Y^U \rightarrow
\left[\begin{array}{ccc}
e^{i \bar{\phi}_{12}} & 0 & 0 \\ 0 & e^{i \bar{\phi}_{22}} & 0 \\ 0 &
0 & 1 \\
\end{array}\right] &
Y^{U} &\left[\begin{array}{ccc} e^{i \bar{\phi}_{22}^U} & 0 & 0 \\ 0 &
1 & 0 \\ 0 & 0 & 1 \\
\end{array}\right], \label{phasegone}
\end{eqnarray}
where we have defined $\bar{\phi}_{22} \equiv \phi_{22}^U -
\phi_{22}^D$ and $\bar{\phi}_{12}  \equiv \phi_{12}' - \phi_{12}$.
This amounts to a phase redefinition of the $(U_R)^c$ and $U_L$
fields.

To diagonalise the real matrices obtained from the above phase
rotations, we first diagonalise the heavy 2 by 2 submatrices, then the
light submatrices as shown below,
\begin{eqnarray}
Y^D \rightarrow
\left[\begin{array}{ccc}
 \tilde{c}_2 & \tilde{s}_2 & 0 \\ - \tilde{s}_2 & \tilde{c}_2 & 0 \\ 0
& 0 & 1 \end{array}\right]
\left[\begin{array}{ccc}
 1& 0 & 0 \\ 0 & \tilde{c}_4 & \tilde{s}_4 \\ 0 & -\tilde{s}_4 &
\tilde{c}_4 \\ \end{array}\right] & Y^D &
\left[\begin{array}{ccc}
1 & 0 & 0 \\ 0 & \bar{c}_4 & -\bar{s}_4 \\ 0 & \bar{s}_4 & \bar{c}_4
\\ \end{array}\right]
\left[\begin{array}{ccc}
c_2 & -s_2 & 0 \\ s_2 & c_2 & 0 \\ 0 & 0 & 1 \\ \end{array}\right]
\nonumber \\ Y^U \rightarrow
\left[\begin{array}{ccc}
 \tilde{c}_1 & \tilde{s}_1 & 0\\ -\tilde{s}_1 & \tilde{c}_1 & 0\\ 0 &
 0 & 1 \end{array}\right]
\left[\begin{array}{ccc}
1 & 0 & 0 \\ 0 & \tilde{c}_3 & \tilde{s}_3 \\ 0 & -\tilde{s}_3 &
\tilde{c}_3 \\ \end{array}\right] & Y^U &
\left[\begin{array}{ccc}
1 & 0 & 0 \\ 0 & \bar{c}_3 & -\bar{s}_3 \\ 0 & \bar{s}_3 & \bar{c}_3
\\ \end{array}\right]
\left[\begin{array}{ccc}
c_1 & -s_1 & 0 \\ s_1 & c_1 & 0 \\ 0 & 0 & 1 \\ \end{array}\right]
\label{diags},
\end{eqnarray}
where $c_i, \bar{s}_i$ refer to $\cos \theta_i$ and $\sin
\bar{\theta}_i$ respectively. Note that since $Y^{U,D}$ are not
symmetric $\tilde{c}_i, \tilde{s}_i$ are independent of $c_i,
\bar{s}_i$.

The diagonal Yukawa couplings of the strange quark and muon obtained
from Eq.\ref{diags} are $(\lambda_s)_{M_X}=H_{22}^D$ and
$(\lambda_\mu)_{M_X} =l H_{22}^D$ since the 22 eigenvalues are just
the 22 elements in this case.  The first family diagonal Yukawa
couplings for the down quark and electron are related by
\begin{equation}
\left( \frac{\lambda_d}{\lambda_e} \right)_{M_X}= l \frac{x_{21}^D
x_{12}^D}{x_{21}^E x_{12}^E}.
\label{givesk}
\end{equation}
We identify the right hand side of Eq.\ref{givesk} with $k$ in
Eq.\ref{1stpred}. The angles are given by $\bar{s}_4=-x_{32}^DH_{32}/
H_{33}$, $s_2=-x_{21}^DH_{21}/(\lambda_s)_{M_X}$,
$s_1=-x_{21}^UH_{21}/(\lambda_c)_{M_X}$ and
$\bar{s}_3=-x^U_{32}H_{32}/H_{33}$.  Note that
$(\lambda_u)_{M_X}=-{x'}_{12}^U x_{21}^UH_{12}' H_{21} /
(\lambda_c)_{M_X}$ is small and in the limit ${O_{12}}' \rightarrow 0$
the up quark is massless in the model.

\begin{figure}
\begin{center}
\leavevmode
\hbox{\epsfxsize=4.5in
\epsfysize=2.6in
\epsffile{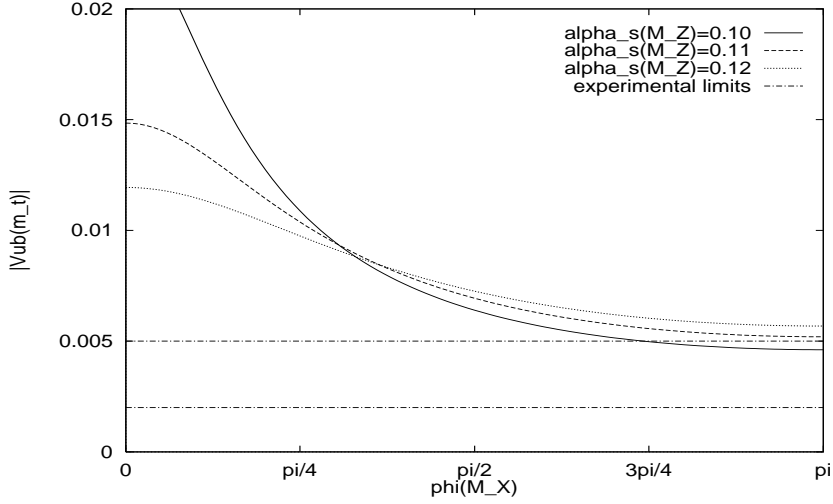}}
\end{center}
\caption{$|V_{ub}|$ bounds predicted in terms of the complex phase
$\phi$ for various $\alpha_s (M_Z)$ and the ansatze $A_{1-6}$,
corresponding to $l=3$.}
\label{Vubpred}
\end{figure}

\begin{figure}
\begin{center}
\leavevmode
\hbox{\epsfxsize=4.5in
\epsfysize=2.6in
\epsffile{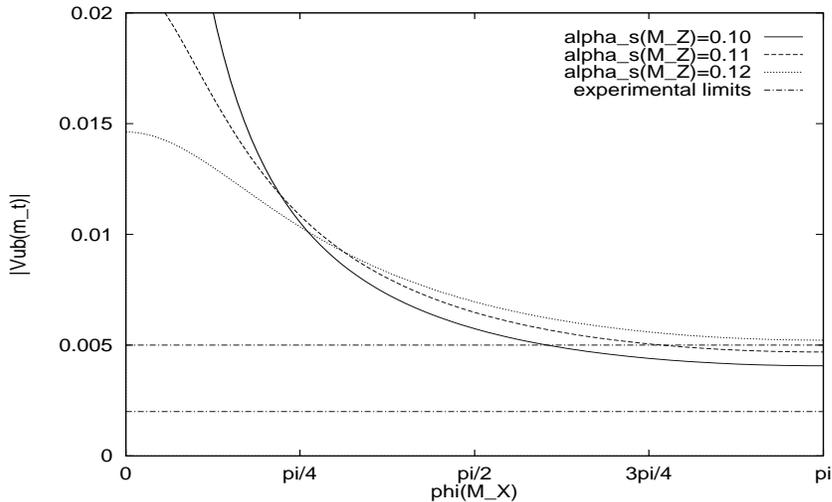}}
\end{center}
\caption{$|V_{ub}|$ bounds predicted in terms of the complex phase
$\phi$ for various $\alpha_s (M_Z)$ and the ansatze $A_{7,8}$,
corresponding to $l=4$.}
\label{Vubpred2}
\end{figure}

Denoting $\theta_3=\bar{\theta}_3-\bar{\theta}_4$ and $\phi = -
\bar{\phi}_{22}$ , we substitute the
diagonalising matrices from Eqs.~\ref{diags} and~\ref{phasegone} into
the CKM matrix in Eq.\ref{CKM} to obtain
\begin{equation}
V_{CKM} = \left[\begin{array}{ccc} c_2 c_1 e^{i \phi} + s_2 s_1 c_3 &
-s_2 c_1 e^{i \phi} + s_1 c_2 c_3 & s_1 s_3 \\ -s_1 c_2 e^{i \phi} +
c_1 s_2 c_3 & s_1 s_2 e^{i \phi} + c_2 c_3 c_1 & s_3 c_1 \\ -s_2 s_3 &
-c_2 s_3 & c_3 \end{array}\right].
\end{equation}
Note that $\tan \theta_1 = \frac{V_{ub}}{V_{cb}}$.  This is a generic
feature of all the ansatze. To obtain a prediction for $|V_{ub}|$ we
note that
\begin{eqnarray}
|V_{us}| &=& |-s_2 c_1 e^{i \phi} + s_1 c_2 c_3| \nonumber \\ &\sim&
         |\frac{V_{ub}}{V_{cb}}| |\frac{x_{21}^D
\lambda_c}{x_{21}^U \lambda_s}e^{i \phi} - 1   | \nonumber \\
\Rightarrow
|V_{ub}(m_t)|&\sim & \frac{ |V_{us}(m_t)| |V_{cb}(m_t)|}{\sqrt{ 1 +
\left( \frac{x_{21}^D \lambda_c(M_X)}{x_{21}^U \lambda_s(M_X)}
\right)^2 - 2 \cos\phi(M_X)
\frac{x_{21}^D \lambda_c(M_X)}{x_{21}^U \lambda_s(M_X)} } }.
\label{Vubpredict}
\end{eqnarray}
Eq.\ref{Vubpredict} predicts a value of $|V_{ub}|$ that is dependent
upon the value of $l$. This is because of the appearance of
$\lambda_s$ in Eq.\ref{Vubpredict}, which is predicted to have
different values in Eq.\ref{2ndpred} depending on $l$. The $|V_{ub}|$
predicted is displayed in Figs.~\ref{Vubpred} and~\ref{Vubpred2} and
fits the phenomenological values of 0.002--0.005 successfully only for
large values of the complex phase. Clearly
Figs.\ref{Vubpred},\ref{Vubpred2} predict large values of $|V_{ub}|$.
We emphasise that this prediction applies to all of the successful
ansatze $A_i, B_i$.

\section{Conclusions}
We have discussed the problem of fermion masses in the supersymmetric
SU(4)$\otimes$ SU(2)$_L \otimes$SU(2)$_R$ model, where the gauge group
is assumed to be broken to the standard gauge group at $M_X \sim
10^{16}$ GeV. Although the gauge group is not unified at $M_X$, it is
hoped that the model may be embedded in some string theory near the
Planck scale. Since the model involves no adjoint representations of
the gauge group, and has no doublet-triplet splitting problem, the
prospects for achieving string unification in this model are very
good, and some attempts in this direction have already been made
\cite{leo2}. However here we have restricted ourselves to the
low-energy effective field theory near the scale $M_X \sim 10^{16}$
GeV, and parameterised the effects of string unification by
non-renormalisable operators whose coefficients are suppressed by
powers of $(M_X/M)$, where $M>M_X$ is some higher scale associated
with string physics.

We have assumed that the heavy third family receives its mass from a
single renormalisable Yukawa coupling in the superpotential. The model
predicts third family Yukawa unification (i.e.\ the top, bottom and
tau Yukawa couplings are all equal at $M_X$) leading to predictions
for the top mass and $\tan \beta$. The other Yukawa couplings result
from the effect of non-renormalisable operators of order
$(M_X/M)^{2n}$, where $n=1$ operators are suitable for the lower 2 by
2 block of the Yukawa matrices, and $n=2,3$ are suitable for the upper
2 by 2 block.  In fact we have classified all possible operators in
this model for $n=1$ and $n=2$.

The successful ansatz $A_i, B_i$ in
Eqs.\ref{A1}-\ref{endans},\ref{suclighti}-\ref{suclightii} involve 8
real parameters $H_{33}$, $H_{32}$, $H_{22}^{U,D}$, $H_{21}$,
$H_{12}$, $H_{12}'$, plus an unremovable phase. With these 8
parameters we can describe the 13 physical quantities $\bar{m}_u$,
$\bar{m}_d$, $\bar{m}_s$, $\bar{m}_c$, $\bar{m}_b$, $\bar{m}_t$,
$\bar{m}_e$, $\bar{m}_\mu$, $\bar{m}_\tau$, $\theta_{1,2,3}$,
$\phi$. Third family Yukawa unification led to a prediction for $m_t
(\mbox{pole}) = 130-200$ GeV and $\tan \beta = 35-65$, depending on
$\alpha_S (M_Z)$ and $\bar{m}_b$. More accurate predictions could be
obtained if the error on $\alpha_S(M_Z)$ and $\bar{m}_b$ were
reduced. The analysis of the lower 2 by 2 block led to 2 possible
predictions for $\bar{m}_s$ depending on whether $l=3$ or $4$, as
shown in Fig.\ref{ms} ($l=3$ is the GJ prediction). In the upper 2 by
2 block analysis we were led to 5 possible predictions for
$\bar{m}_d$, depending on whether $k=2,8/3,3,4,16/3$, as shown in
Fig.\ref{md} (again $k=3$ is the GJ prediction.) Finally, we have 2
predictions for $|V_{ub}|$ depending on whether $l=3$ or 4, as shown
in Figs.\ref{Vubpred},\ref{Vubpred2}. Both predict a large value of
$|V_{ub}|$, depending on $\phi$.

The high values of $\tan \beta$ required by our model (also predicted
in SO(10)) can be arranged by a suitable choice of soft SUSY breaking
parameters as discussed in ref.\cite{Carenaetal}, although this leads
to a moderate fine tuning problem \cite{Yuk}. The high value of $\tan
\beta$ is not stable under radiative corrections unless some other
mechanism such as extra approximate symmetries are invoked.  $m_t$ may
have been overestimated, since for high $\tan \beta$, the equations
for the running of the Yukawa couplings in the MSSM can get
corrections of a significant size from Higgsino--stop and
gluino--sbottom loops. The size of this effect depends upon the mass
spectrum and may be as much as 30 GeV. For our results to be
quantitatively correct, the sparticle corrections to $m_b$ must be
small. This could happen in a scenario with non-universal soft
parameters, for example. Not included in our analysis are threshold
effects, at low or high energies. These could alter our results by
several per cent and so it should be borne in mind that all of the
mass predictions have a significant uncertainty in them. It is also
unclear how reliable 3 loop perturbative QCD at 1 GeV is.

Compared to SO(10) \cite{Larry}, the lack of predictivity of our model
is somewhat discouraging. In the SO(10) model, the spectrum is
described by just 4 operators, whereas in our model the spectrum is
described by 7 operators. One basic reason for this is that, unlike
SO(10), our Yukawa matrices are inherently asymmetric.  In order to
fill out our Yukawa matrices, we need to add operators in the $ij$ and
$ji$ positions separately.  This of course permits asymmetric texture
zeroes such as those in $A_{1-8}$, which have not been studied before.
In the upper 2 by 2 block, the SO(10) model can satisfy all of the
phenomenological constraints with just 2 operators, the up Yukawa
coupling becoming very small through a small Clebsch ratio
$(1/27)^2$. This permits the SO(10) model to make predictions for the
up quark mass and the complex phase as well as $|V_{ub}|$. Whereas we
cannot predict $\bar{m}_u$ in this model, a natural explanation is
given for its relatively small value in terms of a higher dimensional
operator.  Note that simply not having this operator would not alter
any of the predictions, except that the up quark would be massless,
thus solving the strong CP problem. Thus a simple and natural way of
obtaining a massless up quark is given by our model with
$O^{n=3}\rightarrow 0$ in Eqs.~\ref{suclighti}-\ref{suclightii}, which
would reduce the number of operators in our model by one.

Despite the lack of predictivity of the model compared to SO(10), the
SU(4) $\otimes$SU(2)$_L \otimes$SU(2)$_R$ model has the twin
advantages of having no doublet-triplet splitting problem, and
containing no adjoint representations, making the model technically
simpler to embed into a realistic string theory. Although both these
problems can be addressed in the SO(10) model~\cite{GK23split,new}, we
find it encouraging that such problems do not arise in the first place
in the SU(4)$\otimes$SU(2)$_L \otimes$SU(2)$_R$ model.  Of course
there are other models which also share these advantages such as
flipped SU(5) or even the standard model.  However, at the field
theory level, such models do not lead to Yukawa unification, or have
precise Clebsch relations between the operators describing the light
fermion masses.  It is the combination of all of the attractive
features mentioned above which singles out the present model for
serious consideration.

\vspace{\baselineskip}
{\Large {\bf Acknowledgments}}
\vspace{\baselineskip}

B. C. Allanach would like to thank PPARC for financial support in the
duration of this work.
\newpage
{\Large {\bf Appendix 1 : $n=1$ Operators}}
\vspace{\baselineskip}

Following is a list of all $n=1$ non--renormalisable operators.  These
operators are constructed from various group theoretic contractions of
the following five fields,
\begin{equation}
O^{\alpha \rho y w}_{\beta \gamma x z} = F^{\alpha a} \bar{F}_{\beta
x} h^y_a \bar{H}_{\gamma z} H^{\rho w}.
\label{n1ops}
\end{equation}
It is useful to define some SU(4) invariant tensors $C$, and SU(2)$_R$
invariant tensors $R$ as follows:
\begin{eqnarray}
(C_1)^{\alpha}_\beta &=& \delta^\alpha_\beta \nonumber \\
(C_{15})^{\alpha \rho}_{\beta \gamma} &=& \delta^\beta_\gamma
\delta^\rho_\alpha - \frac{1}{4} \delta^\beta_\alpha
\delta^\rho_\gamma \nonumber \\
(C_6)_{\alpha \beta}^{\rho \gamma} &=& \epsilon_{\alpha \beta \omega
\chi} \epsilon^{\rho \gamma \omega \chi} \nonumber \\
(C_{10})^{\alpha \beta}_{\rho \gamma} &=& \delta^\alpha_\rho
\delta^\beta_\gamma + \delta^\alpha_\gamma \delta^\beta_\rho \nonumber
\\
(R_1)^x_y &=& \delta^x_y \nonumber \\ (R_3)^{wx}_{yz} &=& \delta^x_y
\delta^w_z - \frac{1}{2} \delta^x_z
\delta^w_y \nonumber \\
(R_S)^{wx}_{yz} &=& \delta_y^w \delta_z^x + \delta^w_z \delta^x_y ,
\label{T1s}
\end{eqnarray}
where $\delta^\alpha_\beta$, $\epsilon_{\alpha \beta \omega \chi}$,
$\delta^x_y$, $\epsilon_{wz}$ are the usual invariant tensors of
SU(4), SU(2)$_R$. The SU(4) indices on $C_{1,6,10,15}$ are contracted
with the SU(4) indices on two fields to combine them into
$\underline{1}$, $\underline{6}$, $\underline{10}$, $\underline{15}$
representations of SU(4) respectively. Similarly, the SU(2)$_R$
indices on $R_{1,3}$ are contracted with SU(2)$_R$ indices on two of
the fields to combine them into $\underline{1}$, $\underline{3}$
representation of SU(2)$_R$.

The operators in Tables~\ref{table:opclass},\ref{table:opcombs} are
then given explicitly by contracting Eq.\ref{n1ops} with the invariant
tensors of Eq.\ref{T1s} in the following manner:
\begin{eqnarray}
O^A & \sim & (C_1)^\beta_\alpha (C_1)^\gamma_\rho (R_1)^z_w (R_1)^x_y
O^{\alpha \rho y w}_{\beta \gamma x z} \nonumber
\\  O^B & \sim & (C_1)^\beta_\alpha (C_1)^\gamma_\rho (R_3)^{zq}_{wr}
(R_3)^{xr}_{yq} O^{\alpha \rho y w}_{\beta \gamma x z} \nonumber \\
O^C & \sim & (C_{15})^{\beta \chi}_{\alpha \sigma} (C_{15})^{\gamma
\sigma}_{\rho \chi} (R_1)^z_w (R_1)^x_y O^{\alpha \rho y w}_{\beta
\gamma x z}\nonumber \\  O^D & \sim & (C_{15})^{\beta \chi}_{\alpha
\sigma} (C_{15})^{\gamma \sigma}_{\rho \chi} (R_3)^{zq}_{wr}
(R_3)^{xr}_{yq}O^{\alpha \rho y w}_{\beta \gamma x z} \nonumber \\ O^E
& \sim & (C_6)^{\omega \chi}_{\alpha \rho} (C_6)^{\beta
\gamma}_{\omega \chi} \epsilon^{zx} \epsilon_{yw} O^{\alpha \rho y w}_{\beta
\gamma x z} \nonumber \\
O^F & \sim & (C_6)^{\omega \chi}_{\alpha
\rho} (C_6)^{\beta \gamma}_{\omega \chi} (R_3)^{sq}_{wr}
(R_3)^{xr}_{tq} \epsilon_{ys} \epsilon^{zt} O^{\alpha \rho y w}_{\beta
\gamma x z} \nonumber \\
O^G & \sim &(C_{10})^{\omega \chi}_{\alpha \rho} (C_{10})^{\beta
\gamma}_{\omega \chi} \epsilon^{xz}
\epsilon_{yw}O^{\alpha \rho y w}_{\beta \gamma x z} \nonumber \\  O^H
 & \sim & (C_{10})^{\omega \chi}_{\alpha \rho} (C_{10})^{\beta
 \gamma}_{\omega \chi} (R_S)^{xz}_{qr} (R_S)^{qr}_{yw} O^{\alpha \rho
 y w}_{\beta \gamma x z} \nonumber \\ O^I & \sim & (C_1)^\gamma_\alpha
 (C_1)^\beta_\rho (R_1)^z_y (R_1)^x_wO^{\alpha \rho y w}_{\beta \gamma
 x z} \nonumber \\ O^J & \sim & (C_1)^\beta_\rho (C_1)^\gamma_\alpha
 (R_3)^{zq}_{yr} (R_3)^{xr}_{wq}O^{\alpha \rho y w}_{\beta \gamma x z}
\nonumber \\
 O^K & \sim & (C_{15})^{\gamma \omega}_{\alpha \chi} (C_{15})^{\beta
\chi}_{\rho \omega} (R_1)^z_y (R_1)^x_wO^{\alpha \rho y w}_{\beta
\gamma x z} \nonumber \\
 O^L & \sim & (C_{15})^{\gamma \omega}_{\alpha \chi} (C_{15})^{\beta
\chi}_{\rho \omega} (R_3)^{zq}_{yr} (R_3)^{xr}_{wq} O^{\alpha \rho y
w}_{\beta \gamma x z} \nonumber \\ O^M & \sim & (C_1)^\beta_\alpha
 (C_1)^\gamma_\rho \epsilon^{zx}
\epsilon_{wy}O^{\alpha \rho y w}_{\beta \gamma x z} \nonumber \\
 O^N & \sim & (C_6)^{\omega \chi}_{\alpha \rho} (C_6)^{\beta
\gamma}_{\omega \chi}(R_1)^z_y (R_1)^x_wO^{\alpha \rho y w}_{\beta
\gamma x z} \nonumber \\
 O^O & \sim & (C_{10})^{\omega \chi}_{\alpha \rho} (C_{10})^{\beta
\gamma}_{\omega \chi} (R_1)^z_y (R_1)^x_wO^{\alpha \rho y w}_{\beta
\gamma x z} \nonumber \\
 O^P & \sim & (C_{15})^{\gamma \omega}_{\alpha \chi} (C_{15})^{\beta
\chi}_{\rho \omega}(R_1)^z_w (R_1)^x_y O^{\alpha \rho y w}_{\beta
\gamma x z}.
\label{operators}
\end{eqnarray}

\vspace{\baselineskip}
{\Large {\bf Appendix 2: $n=2$ Operators}}
\vspace{\baselineskip}

The $n=2$ operators are formed from the following seven fields:
\begin{equation}
O^{\rho_1 \rho_2 \rho_3 a o q}_{\gamma_1 \gamma_2 \gamma_3 m p n} =
F^{\rho_1 a} \bar{H}_{\gamma_1 m} \bar{F}_{\gamma_2 p} H^{\rho_2 o}
\bar{H}_{\gamma_3 n} H^{\rho_3 q} h^r_a. \label{n2ops}
\end{equation}
Apart from the invariants defined in Eq.\ref{T1s}, we also define
\begin{eqnarray}
(C_{20})^{\alpha \beta \gamma}_{\rho \sigma \omega} &=&
\delta^\alpha_\sigma \delta^\beta_\rho \delta^\gamma_\omega +
\delta^\alpha_\sigma \delta^\beta_\omega \delta^\gamma_\rho +
\delta^\alpha_\rho \delta^\beta_\omega \delta^\gamma_\sigma +
\delta^\alpha_\omega \delta^\beta_\sigma \delta^\gamma_\rho +
\delta^\alpha_\omega \delta^\beta_\rho \delta^\gamma_\sigma +
\delta^\alpha_\rho \delta^\beta_\sigma \delta^\gamma_\omega \nonumber
\\
(R_4)^{m n t}_{b c k} &=&
\delta^m_c \delta^n_b \delta^t_k +
\delta^m_c \delta^n_k \delta^t_b +
\delta^m_b \delta^n_k \delta^t_c +
\delta^m_k \delta^n_c \delta^t_b +
\delta^m_k \delta^n_b \delta^t_c +
\delta^m_b \delta^n_c \delta^t_k,
\end{eqnarray}
where the possible SU(4) structures which multiply Eq.\ref{n2ops} are
\begin{eqnarray}
(A)^{\gamma_1 \gamma_2 \gamma_3}_{\rho_1 \rho_2 \rho_3}
&=&(C_{15})^{\gamma_1 \alpha}_{\rho_1 \beta} (C_{15})^{\gamma_2
\beta}_{\rho_2 \alpha} (C_1)^{\gamma_3}_{\rho_3} \nonumber \\
(B)^{\gamma_1 \gamma_2 \gamma_3}_{\rho_1 \rho_2 \rho_3} &=&
(C_{15})^{\gamma_3 \alpha}_{\rho_3 \beta} (C_{15})^{\gamma_2
\beta}_{\rho_2 \alpha} (C_1)^{\gamma_1}_{\rho_1} \nonumber \\
(C)^{\gamma_1 \gamma_2 \gamma_3}_{\rho_1 \rho_2 \rho_3} &=&
(C_{15})^{\gamma_1 \alpha}_{\rho_1 \beta} (C_{15})^{\gamma_3
\beta}_{\rho_3 \alpha} (C_1)^{\gamma_2}_{\rho_2} \nonumber \\
(D)^{\gamma_1 \gamma_2 \gamma_3}_{\rho_1 \rho_2 \rho_3} &=&
(C_1)^{\gamma_2}_{\rho_1} (C_1)^{\gamma_1}_{\rho_2}
(C_1)^{\gamma_3}_{\rho_3} \nonumber \\ (E)^{\gamma_1 \gamma_2
\gamma_3}_{\rho_1 \rho_2 \rho_3} &=& (C_1)^{\gamma_1}_{\rho_1}
(C_1)^{\gamma_2}_{\rho_2} (C_1)^{\gamma_3}_{\rho_3} \nonumber \\
(F)^{\gamma_1 \gamma_2 \gamma_3}_{\rho_1 \rho_2 \rho_3} &=&
(C_{20})^{\alpha \beta \gamma}_{\rho_1 \rho_2 \rho_3}
(C_{20})^{\gamma_1 \gamma_2 \gamma_3}_{\alpha \beta \gamma}
\nonumber \\
(G)^{\gamma_1 \gamma_2 \gamma_3}_{\rho_1 \rho_2 \rho_3} &=&
(C_{10})^{\alpha \beta}_{\rho_1 \rho_2} \epsilon_{\alpha
\rho_3 \mu \sigma} (C_{10})^{\gamma_1 \gamma_2}_{\omega \beta}
\epsilon^{\omega \gamma_3 \mu \sigma}\nonumber \\
(H)^{\gamma_1 \gamma_2 \gamma_3}_{\rho_1 \rho_2 \rho_3}
&=&(C_{10})^{\alpha \beta}_{\rho_3 \rho_2} \epsilon_{\alpha
\rho_1 \mu \sigma} (C_{10})^{\gamma_1 \gamma_2}_{\omega \beta}
\epsilon^{\omega \gamma_3 \mu \sigma}\nonumber \\
(I)^{\gamma_1 \gamma_2 \gamma_3}_{\rho_1 \rho_2 \rho_3} &=&
(C_{10})^{\alpha \beta}_{\rho_1 \rho_3} \epsilon_{\alpha
\rho_2 \mu \sigma} (C_{10})^{\gamma_1 \gamma_2}_{\omega \beta}
\epsilon^{\omega \gamma_3 \mu \sigma}\nonumber \\
(J)^{\gamma_1 \gamma_2 \gamma_3}_{\rho_1 \rho_2 \rho_3} &=&
(C_{10})^{\alpha \beta}_{\rho_1 \rho_2} \epsilon_{\alpha
\rho_3 \mu \sigma} (C_{10})^{\gamma_1 \gamma_3}_{\omega \beta}
\epsilon^{\omega \gamma_2 \mu \sigma}\nonumber \\
(K)^{\gamma_1 \gamma_2 \gamma_3}_{\rho_1 \rho_2 \rho_3} &=&
(C_{10})^{\alpha \beta}_{\rho_3 \rho_2} \epsilon_{\alpha
\rho_1 \mu \sigma} (C_{10})^{\gamma_1 \gamma_3}_{\omega \beta}
\epsilon^{\omega \gamma_2 \mu \sigma}\nonumber \\
(L)^{\gamma_1 \gamma_2 \gamma_3}_{\rho_1 \rho_2 \rho_3} &=&
(C_{10})^{\alpha \beta}_{\rho_1 \rho_3} \epsilon_{\alpha
\rho_2 \mu \sigma} (C_{10})^{\gamma_1 \gamma_3}_{\omega \beta}
\epsilon^{\omega \gamma_2 \mu \sigma}\nonumber \\
(M)^{\gamma_1 \gamma_2 \gamma_3}_{\rho_1 \rho_2 \rho_3} &=&
(C_{10})^{\alpha \beta}_{\rho_1 \rho_2} (C_{10})^{\gamma_1
\gamma_2}_{\alpha \beta} (C_1)^{\gamma_3}_{\rho_3} \nonumber \\
(N)^{\gamma_1 \gamma_2 \gamma_3}_{\rho_1 \rho_2 \rho_3} &=&
(C_{10})^{\alpha \beta}_{\rho_3 \rho_2} (C_{10})^{\gamma_1
\gamma_2}_{\alpha \beta} (C_1)^{\gamma_3}_{\rho_1} \nonumber \\
(O)^{\gamma_1 \gamma_2 \gamma_3}_{\rho_1 \rho_2 \rho_3} &=&
(C_{10})^{\alpha \beta}_{\rho_1 \rho_2} (C_{10})^{\gamma_3
\gamma_2}_{\alpha \beta} (C_1)^{\gamma_1}_{\rho_3} \nonumber \\
(P)^{\gamma_1 \gamma_2 \gamma_3}_{\rho_1 \rho_2 \rho_3} &=&
(C_{10})^{\alpha \beta}_{\rho_3 \rho_2} (C_{10})^{\gamma_3
\gamma_2}_{\alpha \beta} (C_1)^{\gamma_1}_{\rho_1} \nonumber \\
(Q)^{\gamma_1 \gamma_2 \gamma_3}_{\rho_1 \rho_2 \rho_3} &=&
\epsilon_{\rho_1 \rho_2 \alpha \beta} \epsilon^{\gamma_1
\gamma_2 \alpha \beta} (C_1)^{\gamma_3}_{\rho_3} \nonumber \\
(R)^{\gamma_1 \gamma_2 \gamma_3}_{\rho_1 \rho_2 \rho_3} &=&
\epsilon_{\rho_3 \rho_2 \alpha \beta} \epsilon^{\gamma_1
\gamma_2 \alpha \beta} (C_1)^{\gamma_3}_{\rho_1} \nonumber \\
(S)^{\gamma_1 \gamma_2 \gamma_3}_{\rho_1 \rho_2 \rho_3} &=&
\epsilon_{\rho_1 \rho_2 \alpha \beta} \epsilon^{\gamma_1
\gamma_3 \alpha \beta} (C_1)^{\gamma_2}_{\rho_3} \nonumber \\
(T)^{\gamma_1 \gamma_2 \gamma_3}_{\rho_1 \rho_2 \rho_3} &=&
\epsilon_{\rho_3 \rho_2 \alpha \beta} \epsilon^{\gamma_1
\gamma_3 \alpha \beta} (C_1)^{\gamma_2}_{\rho_1} \nonumber.
\end{eqnarray}
The SU(2)$_R$ structures which multiply Eq.\ref{n2ops} are
\begin{eqnarray}
(a)^{mnp}_{oqr} &=& (R_4)^{mnt}_{bck} (R_4)^{bck}_{sqr} \epsilon^{sp}
\epsilon_{to} \nonumber \\
(b)^{mnp}_{oqr} &=& (R_4)^{bck}_{toq} (R_4)^{pns}_{bck} \epsilon^{mt}
\epsilon_{sr} \nonumber \\
(c)^{mnp}_{oqr} &=& (R_4)^{bck}_{oqr} (R_4)^{pmn}_{bck} \nonumber \\
(d)^{mnp}_{oqr} &=& (R_4)^{mns}_{bck} (R_4)^{bck}_{toq} \epsilon_{rs}
\epsilon^{tp} \nonumber \\
(e)^{mnp}_{oqr} &=& (R_4)^{bck}_{tor} (R_4)^{pns}_{bck} \epsilon^{tm}
\epsilon_{sq} \nonumber \\
(f)^{mnp}_{oqr} &=& (R_1)^p_r (R_1)^m_o (R_1)^n_q \nonumber \\
(g)^{mnp}_{oqr} &=& (R_1)^p_o (R_1)^m_r (R_1)^n_q \nonumber \\
(h)^{mnp}_{oqr} &=& (R_1)^n_q \epsilon^{mp} \epsilon_{ro} \nonumber \\
(i)^{mnp}_{oqr} &=& (R_1)^n_r \epsilon^{mp} \epsilon_{qo} \nonumber \\
(j)^{mnp}_{oqr} &=& (R_3)^{ps}_{rt} (R_3)^{mt}_{os} (R_1)^n_q
\nonumber \\
(k)^{mnp}_{oqr} &=& (R_3)^{ns}_{qt} (R_3)^{mt}_{os} (R_1)^p_r
\nonumber \\
(l)^{mnp}_{oqr} &=& (R_3)^{ps}_{ot} (R_3)^{mt}_{rs} (R_1)^n_q
\nonumber \\
(m)^{mnp}_{oqr} &=& (R_3)^{ps}_{ot} (R_3)^{nt}_{qs} (R_1)^m_r
\nonumber \\
(n)^{mnp}_{oqr} &=& (R_3)^{ms}_{rt} (R_3)^{nt}_{qs} (R_1)^p_o
\nonumber \\
(o)^{mnp}_{oqr} &=& (R_1)^p_s (R_3)^{tx}_{ry} (R_3)^{ny}_{qx}
\epsilon_{to}
\epsilon^{sm}\nonumber \\
(p)^{mnp}_{oqr} &=& (R_1)^n_q (R_3)^{tx}_{ry} (R_3)^{ny}_{qx}
\epsilon_{to}
\epsilon^{sm} \nonumber \\
(q)^{mnp}_{oqr} &=& (R_1)^t_r (R_3)^{nx}_{qy} (R_3)^{py}_{sx}
\epsilon_{to}
\epsilon^{sm} \nonumber \\
(r)^{mnp}_{oqr} &=& (R_1)^p_t (R_3)^{nx}_{ry} (R_3)^{sy}_{ox}
\epsilon^{tm}
\epsilon_{sq} \nonumber \\
(s)^{mnp}_{oqr} &=& (R_1)^n_r (R_3)^{px}_{ty} (R_3)^{sy}_{ox}
\epsilon^{tm}
\epsilon_{sq} \nonumber \\
(t)^{mnp}_{oqr} &=& (R_1)^s_o (R_3)^{nx}_{ry} (R_3)^{py}_{tx}
\epsilon^{tm} \epsilon_{sq}.
\end{eqnarray}
The resulting 400 $n=2$ operators are of the form
\begin{equation}
O^{\Xi \delta} = (\Xi)_{\rho_1 \rho_2 \rho_3}^{\gamma_1 \gamma_2
\gamma_3}
(\delta)^{mpn}_{aoq} O^{\rho_1 \rho_2 \rho_3 a o q}_{\gamma_1
\gamma_2 \gamma_3 m p n},
\end{equation}
where $\Xi = A, B \ldots T$ and $\delta = a,b \ldots t$.

 \end{document}